\documentclass[%
 aip,amsmath,amssymb,
 reprint,%
]{revtex4-1}

\usepackage{graphicx}
\usepackage{dcolumn}
\usepackage{bm}
\usepackage[utf8]{inputenc}
\usepackage[T1]{fontenc}
\usepackage{mathptmx}
\usepackage{color}

\begin{document}

\preprint{AIP/123-QED}

\title[Particles in turbulent separated flow over a bump]{Particles in turbulent separated flow over a bump: effect of the Stokes number and lift force}

\author{J.-P. Mollicone}
\email{j.mollicone@imperial.ac.uk}
\affiliation{Department of Civil and Environmental Engineering, Imperial College London, United Kingdom}
\author{M. Sharifi}
\affiliation{Faculty of Sciences and Engineering, Sorbonne University, Paris, France}
\author{F. Battista}
\affiliation{ENEA C.R. Casaccia, S.M. di Galeria, Rome, Italy}
\author{P. Gualtieri}
\author{C.M. Casciola}
\affiliation{Department of Mechanical and Aerospace Engineering, \\ Sapienza University of Rome, Italy}

\date{\today}

\begin{abstract}
Particle-laden turbulent flow that separates due to a bump inside a channel is simulated to analyse the 
effects of the Stokes number and the lift force on the particle spatial distribution. The fluid friction 
Reynolds number is approximately 900 over the bump, the highest achieved for similar computational 
domains. A range of particle Stokes numbers are considered, each simulated with and without the lift 
force in the particle dynamic equation. When the lift force is included a significant difference in the particle concentration, in the order of thousands, is observed in comparison with cases where the lift force is omitted.
The greatest deviation is in regions of high vorticity, particularly 
at the walls and in the shear layer but results show that the concentration also changes in the bulk of the 
flow away from the walls. The particle behaviour changes drastically when the Stokes number is varied. 
As the Stokes number increases, particles bypass the recirculating region that is formed after the bump 
and their redistribution is mostly affected by the strong shear layer. Particles segregate at the walls and 
particularly accumulate in secondary recirculating regions behind the bump. At higher Stokes numbers, 
the particles create reflection layers of high concentration due to their inertia as they are diverted by the 
bump. The fluid flow is less influential and this enables the particles to enter the recirculating region by 
rebounding off walls and create a focus of high particle concentration.
\end{abstract}

\maketitle

\section{\label{sec:intro} Introduction\protect\\ }

Turbulent flows laden with particles are common in natural phenomena and 
engineering applications. The understanding of turbulence and multiphase flows is considered a 
challenge in both experiments and numerical simulations~\cite{balachandar_2010,elghobashi2019direct}. 
As for single-phase flow simulations, different formulations for the carrier phase in multiphase flows 
can be used such as direct numerical simulation (DNS), the method used here, large-eddy simulation (LES), \cite{marchioli2017large,innocenti2016lagrangian,park2017simple}, 
or Reynolds-averaged Navier-Stokes equations (RANS), \cite{minier2014guidelines,sajjadi2017lattice,Vahidifar2018}.

To these possible descriptions of turbulent flow, Lagrangian methods can be used to couple particles and 
fluid at different levels, i.e.  one-way, two-way or four-way coupling regime,~\cite{elghobashi1994predicting}.
In the one-way coupling the particle volume fraction and mass loading are low, the particles are transported 
by the carrier phase which is not modified.
For higher mass loading and small volume fraction, two-way coupling considers the flow modulation due to the particles. In four-way coupling, collisions and hydrodynamic interaction between particles are significant.

In simulations involving a large number of particle smaller than the Kolmogorov scale, a suitable 
numerical approach is the Lagrangian point-particle method, \cite{toschi_2009,kuerten2016point}. 
In such mixed Eulerian-Lagrangian method, the continuous phase is described in a Eulerian framework whilst the 
dispersed phase by a Lagrangian approach solving the dynamic equation for each particle. 
Particles may be considered to be spherical or, requiring more complex modelling, 
non-spherical particles may be considered~\cite{gustavsson2017statistical,voth2017anisotropic}.

The study of particle-laden turbulent flow has been of interest for decades  by considering 
isotropic turbulence, \cite{squires1991preferential,bragg2015mechanisms}, and homogeneous shear 
turbulence, \cite{nicolai2014inertial,battista2018application}, and cases where the flow is 
confined by solid boundaries to investigate, for example, particle behaviour in turbulent boundary layers, 
\cite{sardina2011large,sardina2012self,li2016direct} or in Couette flow~\cite{bernardini2013effect}. 
Various studies are conducted in wall-bounded flows, \cite{marchioli2008statistics}, and they include 
confined flows such as pipe flows, \cite{marchioli2003direct,picano2009spatial} and channels, 
\cite{sardina2012wall,kulick1994particle}, sometimes with the addition of roughness on the 
surface of the channel's walls to modulate the flow and hence the particle dynamics, 
\cite{liu2016turbulence,de2016interaction,vreman2015turbulence}. 
The study of particle-laden fluid jets is also active due to their vast use in engineering and their occurrence 
in nature, \cite{picano2011dynamics,gualtieri2017turbulence,battista2011intermittent,wu2017particle}. 
Some studies focus, for example, on the effect of the Stokes number on particle behaviour in 
turbulent jets, both experimentally ~\cite{lau2014influence,lau2016effect} and numerically~\cite{wang2017direct}.

In many applications though, the geometry involved is more complex than these standard domains. 
For example, microparticles are used in inhalable drug delivery systems since they provide a non-invasive 
treatment and localised delivery method. Some authors~\cite{abdelaziz2017inhalable} discuss their use in the treatment of lung cancer whilst other~\cite{ni2017nanocrystals} show how microparticles can be used to 
deliver embedded nanocrystals to the lungs. These applications call for numerical simulations to study 
microparticle behaviour in realistic models of human airways~\cite{ghahramani2017numerical, stylianou2016direct} 
together with the effect of different breathing conditions~\cite{rahimi2015cfd}. 
Another health related example is the obstruction in blood vessels due to atherosclerosis which is nowadays also 
studied with the aid of computational modelling~\cite{thondapu2016biomechanical,choi2018flow}.

A geometrical change may be intentional, for example, to enhance mixing of fluid and particles or 
to separate particles from fluid for filtration. Specific geometries can be used for the preferential separation 
of particles when populations of particles with different characteristics are present in a carrier phase. 
In Refs.~\cite{huang_2010,huang_2012} the authors study particulate dispersion and mixing through DNS in a serpentine channel 
by considering a large range of particle Stokes numbers. The authors observe high concentrations of particles 
near the surface of the outer wall and show how the heaviest particles reflect from the wall to form reflection layers 
whilst the lighter particles concentrate in the streaks at the wall. 
A recent study~\cite{noorani_2015} also observe particle reflections in the outer bend of turbulent curved pipe flow 
laden with micro-sized inertial particles. The authors document the modification of particle axial and wall-normal 
velocities and the increase in particle turbulent kinetic energy. 
Other studies~\cite{ault2016vortex} show the importance of geometry for particle-capture mechanisms in branching 
junctions by comparing experiments and numerical simulations. The authors show that the capture is 
dependent on vortex breakdown, a result of the creation and evolution of recirculating regions in the 
system and a crucial factor that determines whether particle accumulation is maximised or eliminated.

The aim of the present paper is to study the microparticles behaviour in a turbulent channel flow with bump at one 
of the walls by means of a Lagrangian point-particle method in an incompressible flow simulated using DNS.
The bump makes the flow separate, creating a strong shear layer and recirculating region, 
\cite{stella2017scaling,krankdirect,schiavo2017turbulent}.
The configuration is nonetheless still accessible to classical statistical tools and turbulence theory for 
the detailed study of turbulence dynamics, \cite{mollicone2018turbulence,Mollicone_2017,passaggia2018optimal,Kahler_2016}. 
To the best of our knowledge, this is the first simulation of such a configuration laden with particles
at a friction Reynolds number of 900 over the bump. 
A  wide range of populations ranging from almost tracers to ballistic particles are addressed.
Additionally, due to the high vorticity present in some regions of the flow, we investigate if and where the lift 
force significantly influences the particle dynamics.

The paper is divided as follows: the simulation setup is described in section \ref{sec:intro}, the results are discussed 
in the \ref{sec:results} and the final remarks are in section \ref{sec:final}.

\section{Simulation Setup}
\label{sec:intro} 

\subsection{Fluid phase}

The computational domain has dimensions $(L_x \times L_y \times L_z) = (26 \times 2 \times 2\pi ) \times h_0 $, 
where $x$, $y$, and $z$ are the streamwise, wall-normal and spanwise coordinates respectively and $h_0$ is 
half the nominal channel height, see figure~\ref{fig:3D_vel}. Periodic boundary conditions are enforced in 
both $x$ and $z$ directions, whilst at the walls no-slip conditions are applied. The bottom wall contains 
a bump that is described by the easily reproducible and differentiable equation (especially important 
for particle rebound) $y=a(1+\cos((2\pi/c)(x-b)))$ where $a=0.25$, $b=3$, $c=2$ and $x$ ranges from $2$ to
$4$. The periodicity in the streamwise direction avoids artificial inflow/outflow boundary conditions and 
the period is chosen as large as possible, within computational limitations, to allow the analysis of an almost 
isolated bump, with definite flow reattachment and negligible streamwise correlation. 
The incoming flow accelerates at the channel restriction and a recirculating region forms behind the bump, 
starting downstream of the bump tip. An intense shear layer separates the recirculating region from the 
outer flow. Downstream of the bump, the flow re-attaches completely. The turbulence dynamics for such 
flows over a bump is discussed in detail in~\cite{Mollicone_2017,mollicone2018turbulence}. 

Direct numerical simulation (DNS) is used to solve the incompressible Navier-Stokes equations,
\begin{equation}
\frac{\partial {\bf u}}{\partial t}+{{\bf u} \cdot \nabla {\bf u} }
=-\nabla p + {\nu \nabla^{2}{\bf u}}   \quad \quad   \nabla \cdot {{\bf u}} = 0  \, ,
\end{equation}
where $\bf{u}$ is the fluid velocity, $t$ is the time, $p$ is the hydrodynamic pressure and $\nu$ is 
the kinematic viscosity. Nek5000~\cite{nek5000}, which is based on the spectral element method (SEM)~\cite{Patera_1984}, is used to solve both the flow domain and the dispersed phase. 
The implemented algorithm for the computation of the particle dynamics is deeply described in 
subsection~\ref{subsec:sf}. The simulations are carried out at bulk Reynolds number 
${\rm Re_0}= h_0 U_b/\nu=10000$, where $U_b$ is the bulk velocity. All length scales are made dimensionless 
with the nominal channel half-height $h_0$, time with $h_0/U_b$ and pressure with $\rho U_b^2$. The 
maximum friction Reynolds number, achieved close to the bump tip, is ${\rm Re}_{\tau} = 900$, defined as 
${\rm Re}_{\tau} =u_\tau h_0/\nu$, where the friction velocity is $u_\tau=\sqrt{\tau_w/\rho}$, $\tau_w$ is the 
local mean shear stress and $\rho$ is the constant fluid density.
The simulation has been performed with about $400$ million grid point on $32768$ cores using approximately 
$30$ million core hours, using spectral elements of order of $N=11$. In this case the grid spacing, $\Delta x^+=6.5$,
$\Delta z^+=7.0$ and $\Delta y^+_{max/min}=9.5/0.9$, is adeguate for the high fidelity description of
all the flow scales. For detailed discussion about the resolution the reader can refer to section 2 of Ref.~\citet{Mollicone_2017}.
Approximately 500 statistically uncorrelated fields, separated by a time interval of $\Delta t_{stat} = 6$,
have been collected for each simulation in order to obtain properly converging statistics. Defining the 
'flow-through time', $t_{ft}$ , as the time needed for a turbulent structure to travel all along the channel length, 
the simulation time is $T_{tot} = 3000 \simeq 115 t_{ft}$ , which makes sure that statistics converge.

\subsection{Solid phase}
\label{subsec:sf}

The solid phase is composed of spherical particles with radius smaller than the dissipative 
scale, the wall unit in our case. In a dilute suspension at low mass loading, the turbulence modulation due to the particles, the inter-particle collisions and the hydrodynamic interactions can be neglected~\cite{balachandar_2010}. 
Under these conditions, the one-way coupling regime can be assumed and the Newton equation is forced 
by the Stokes drag together with the lift force which we intentionally include or omit to investigate its effect, 
namely
\begin{equation}
\frac{d { {\bf x}_p}}{dt}= {\bf v}_p \qquad\frac{d { {\bf v}_p}}{dt}= \underbrace{ \frac{1}{\tau_p}\left({\bf u}\big|_p - {\bf v}_p \right)  }_\text{Stokes Drag} +  
\underbrace{\beta\left({\bf u}\big|_p - {\bf v}_p\right) \times \mbox{\boldmath $\zeta$}\big|_p}_\text{Lift}
\label{eq:part_dyn_wl}
\end{equation}
where ${\bf x}_p$ is the particle position, ${\bf v}_p$ is the particle velocity, ${\bf u}_p$ is the fluid velocity at the particle position and $\tau_p=\left({\rho_p}/{\rho_f} \right)\left({d_p^2}/{18\nu}\right)$ is the particle relaxation time,  
$ \rho_p / \rho_f = 1500$ is the particle to fluid density ratio, $d_p$ is the particle diameter, the coefficient
of the lift force is $\beta=1/\left(2\left( \rho_p/\rho_f\right)\right)$ and \mbox{\boldmath $\zeta$} is the vorticity. 
The particles elastically bounce at the wall, where in correspondence of the bump the exact bouncing direction is 
evaluated by the analytical function of the wall profile.The dynamics of the particle in the one-way coupling regime is well described by the Stokes number, i.e. the ratio between the particle relaxation time and the fluid characteristic time scale. Two different Stokes numbers are defined, the reference one, 
${\rm St}_0 = \tau_p U_b/h_0$, and the viscous Stokes number, ${\rm St}_+= \tau_p u_\tau^2/\nu$. The two are related through 
the Reynolds number, ${\rm St}_0= {\rm St}_+ {\rm Re}_0/{\rm Re}_\tau^2$. 

Equations \eqref{eq:part_dyn_wl} are evolved for each single particle with a fourth order Adams-Bashforth method in time.  A spectral interpolation, intrinsic to the Nek5000 code, is employed to evaluate the fluid variables at the particle position. Twenty different populations are evolved: ten different Stokes numbers $St_+=[0.1, 1, 2, 5, 10, 50, 100, 200, 400, 600]$ each with and without the lift term. Ten million of particle are considered in the simulation. 
The huge number of particles is employed to obtain an adeguate statistical convergence, since each single 
particle does not feel the presence of the other and does not modify the carrier flow. 
The effect of gravity is negligible for the particle population considered in the present study. 
In fact, the dimensionless terminal velocity, $V_t=\tau_p \, g$, can be expressed in terms of the control 
parameters as $V_t/U_b={\rm St}_+ \, {\rm Re}_0/\left( {\rm Fr_0} \, {\rm Re}_{\tau}\right)^2$ where 
${\rm Fr}_0=U_b/\sqrt{g \, h_0}$ is the bulk Froude number. For the heaviest population at ${\rm St}_+=600$, 
it turns out that the ratio $V_t/U_b$ is of the order of $10^{-2}$.


\section{Results}
\label{sec:results}

\begin{figure}
\centering{
\includegraphics[width=0.6\linewidth] {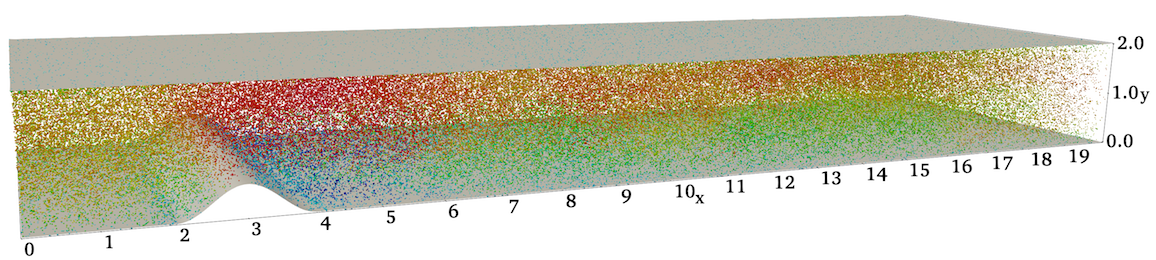}
{\scriptsize \put(-283,57){\bf (a)}}
\includegraphics[width=0.39\linewidth]{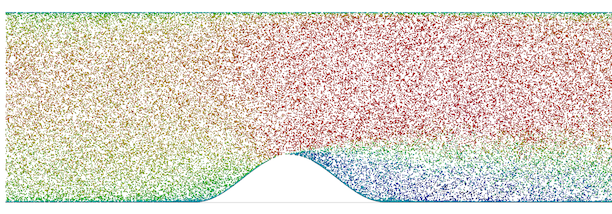}
{\scriptsize \put(-190,57){\bf (b)}}\\
\includegraphics[width=0.6\linewidth] {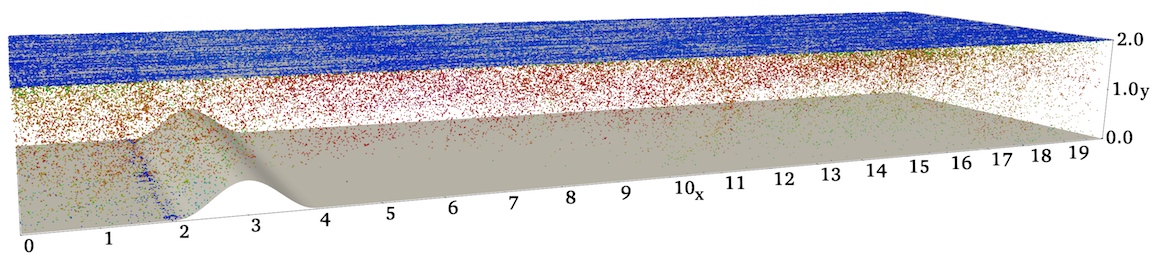}
{\scriptsize \put(-283,57){\bf (c)}}
\includegraphics[width=0.39\linewidth]{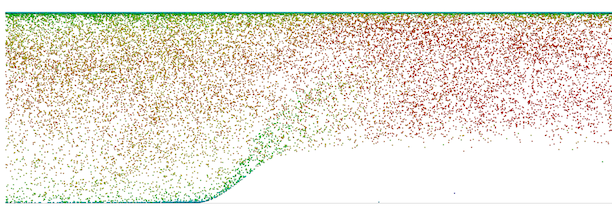}
{\scriptsize \put(-190,57){\bf (d)}}\\	
\includegraphics[width=0.6\linewidth] {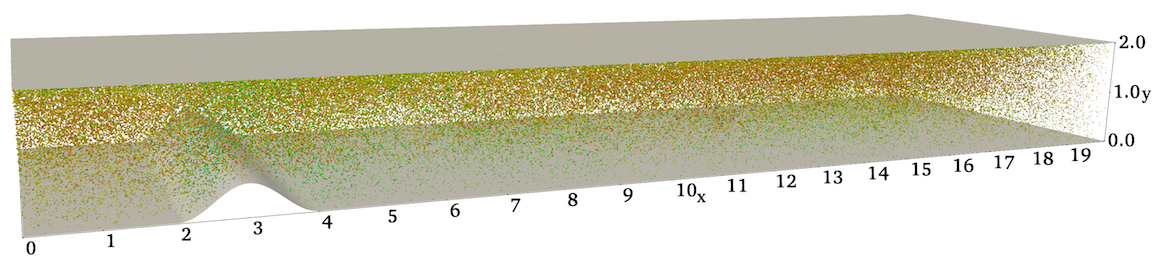}
{\scriptsize \put(-283,57){\bf (e)}}
\includegraphics[width=0.39\linewidth]{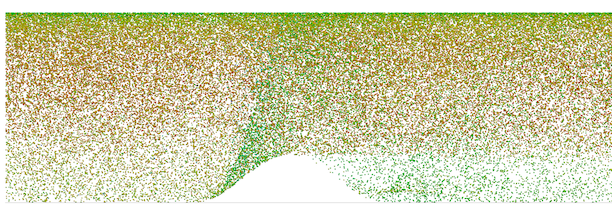}
{\scriptsize \put(-190,57){\bf (f)}}\\
\flushright \includegraphics[width=0.35\linewidth] {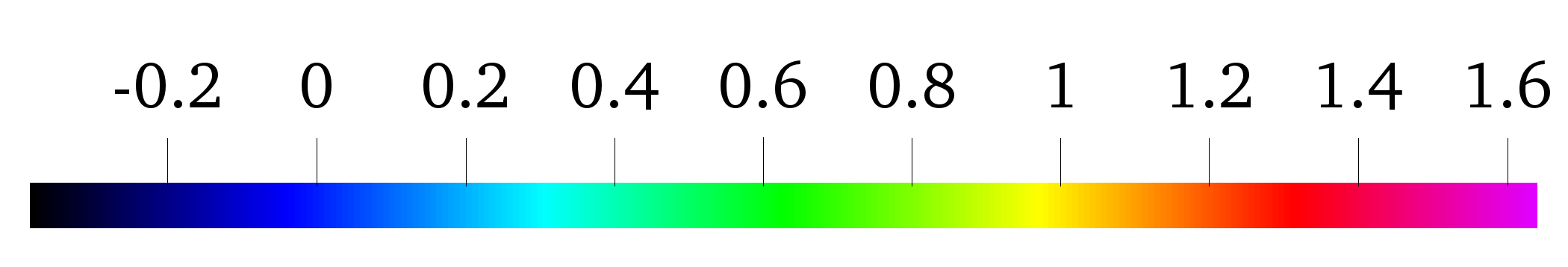}\\  }
\caption{\label{fig:3D_vel} Instantaneous snapshot of particles coloured by stream-wise velocity for 
$St^+= 1$ in panels (a) and (b), $St^+= 50$ in panels (c) and (d), $St^+= 600$ in panels (e) and (f). 
Particles are not to scale.}
\end{figure}
Figure~\ref{fig:3D_vel} shows instantaneous snapshots of the particles coloured with their instantaneous streamwise 
velocity for $St_+=[1, 50, 600]$. The whole computational domain is shown in the left panels whilst the area around the 
bump, in a view orthogonal to the $x$-$y$ plane, is shown in the right panels. Unless otherwise stated, figures show 
the particles that have been evolved with the lift term included in equation~\eqref{eq:part_dyn_wl}. 
At low Stokes number, the particles distribute themselves evenly throughout the domain and have velocities 
comparable to the fluid velocity.
The low particle velocity inside the recirculating region behind the bump contrasts the 
high particle velocity at the centre of the channel. On the other hand, at $St_+=50$, the number of particles in the 
recirculating region is negligible. The particles move towards the upper part of the channel as they are projected upwards 
as they hit the bump. At higher Stokes number, the particles have a ballistic motion and after hitting the bump continue to 
bounce off the upper wall and therefore manage to enter the recirculating region. Particles that do not hit the bump continue 
straight in their trajectory, as shown by the particles above $y=0.5$ behind the bump.

\begin{figure}
\centering{
\includegraphics[width=\linewidth] {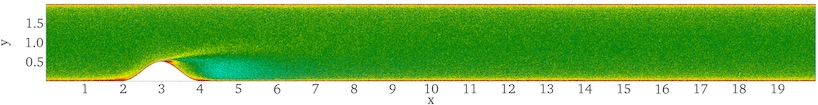}    
{\scriptsize \put(-469,54){\bf (a)}}\\
\includegraphics[width=\linewidth] {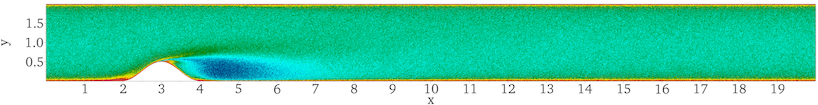}     
{\scriptsize \put(-469,54){\bf (b)}}\\
\includegraphics[width=\linewidth] {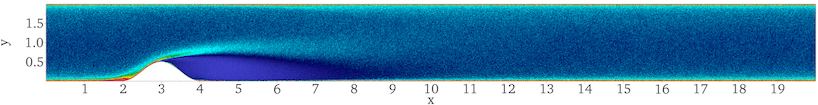}     
{\scriptsize \put(-469,54){\bf (c)}}\\
\includegraphics[width=\linewidth] {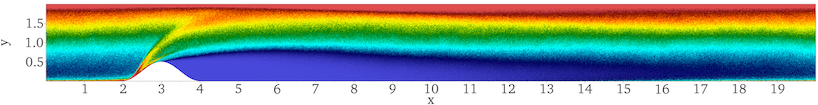}     
{\scriptsize \put(-469,54){\bf (d)}}\\
\includegraphics[width=\linewidth] {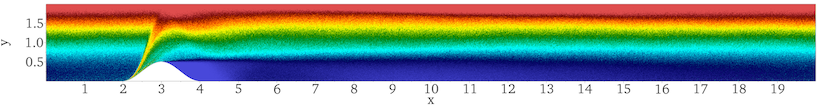}     
{\scriptsize \put(-469,54){\bf (e)}}\\
\includegraphics[width=\linewidth] {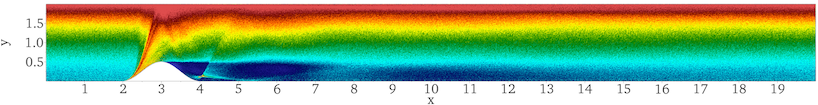}     
{\scriptsize \put(-469,54){\bf (f)}}\\
\flushright \includegraphics[width=0.18\linewidth] {./FIGURES/conc_legend} \\ }
\caption{Mean particle concentration in the $(x,y)$ plane for $St_+=[1, 2, 5, 50, 200, 600]$ in panels (a) to (f) respectively.}
\label{fig:conc}
\end{figure}
\begin{figure}
\centering{
\includegraphics[width=0.49\linewidth] {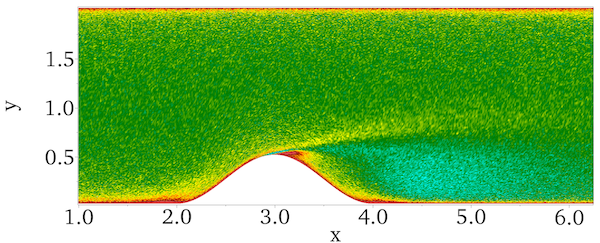}
{\scriptsize \put(-228,86){\bf (a)}}
\includegraphics[width=0.49\linewidth] {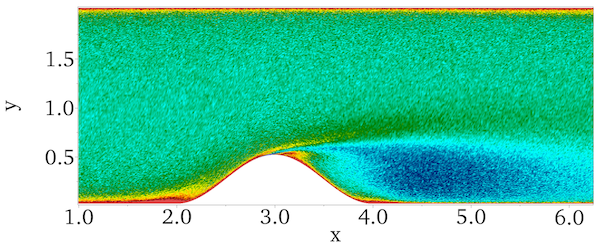}
{\scriptsize \put(-228,86){\bf (b)}}
\includegraphics[width=0.49\linewidth] {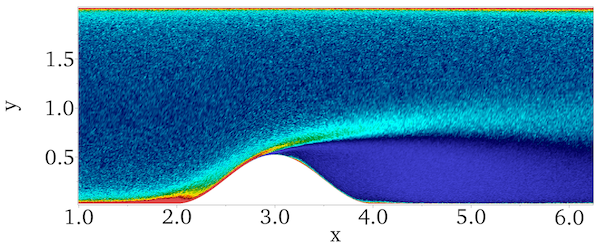}
{\scriptsize \put(-228,86){\bf (c)}}
\includegraphics[width=0.49\linewidth] {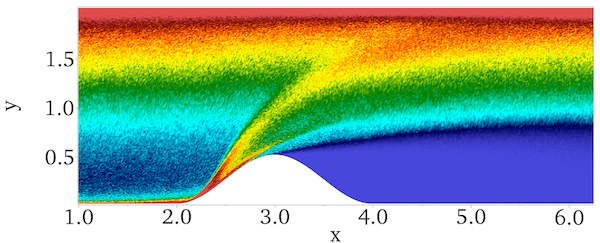}
{\scriptsize \put(-228,86){\bf (d)}}
\includegraphics[width=0.49\linewidth] {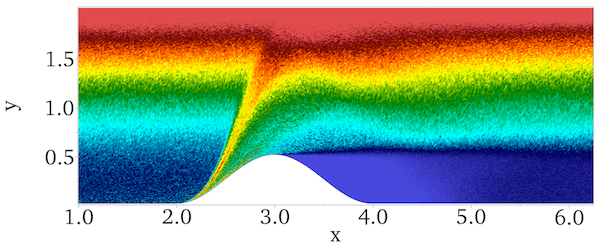}
{\scriptsize \put(-228,86){\bf (e)}}
\includegraphics[width=0.49\linewidth] {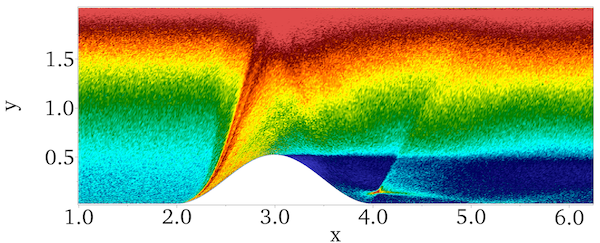}
{\scriptsize \put(-228,86){\bf (f)}}}\\
\flushright \includegraphics[width=0.18\linewidth] {./FIGURES/conc_legend}\\
\caption{ Mean particle concentration around the bump in the $(x,y)$ plane for 
$St_+=[1, 2, 5, 50, 200, 600]$ in panels (a) to (f) respectively.}
\label{fig:conc_zoom}
\end{figure}

The particle behaviour is discussed in detail by considering the mean particle concentration, 
$C(x,y)= \langle \left( n(x,y,z,t) / V(x,y,z) \right) / \left( N_T / V_T \right)\rangle$, where $n(x,y,z,t)$ is the 
instantaneous local number of particles in a cell at $(x,y,z)$ position, $V(x,y,z)$ is the corresponding cell 
volume, $N_T$ is the total number of particles and 
$V_T$ is the total domain volume. The angular brackets denote averaging in the homogeneous 
spanwise direction, $z$, and in time. The normalisation by $N_T/V_T$ represents the homogeneous particle 
concentration, i.e. the concentration value that would be obtained at any point if all the particles were equally
distributed in all the domain. The data sets are taken after the system is statistically stationary and at time
intervals larger than the correlation time. Six out of the total ten particle populations that have been
simulated, $St_+=[1, 2, 5, 50, 200, 600]$, will be presented since the populations that exhibit similar behaviour 
are omitted. The population at $St_+=0.1$ is omitted since particles act as tracers and fill the whole domain 
homogeneously with no effect of the bump and only a slight accumulation at the walls. Figures \ref{fig:conc} 
and \ref{fig:conc_zoom} show the mean particle concentration as a coloured contour plot, with the latter 
zoomed on the region around the bump since it is the area of interest. 

\begin{figure}
\centering{
\includegraphics[width=0.32\linewidth] {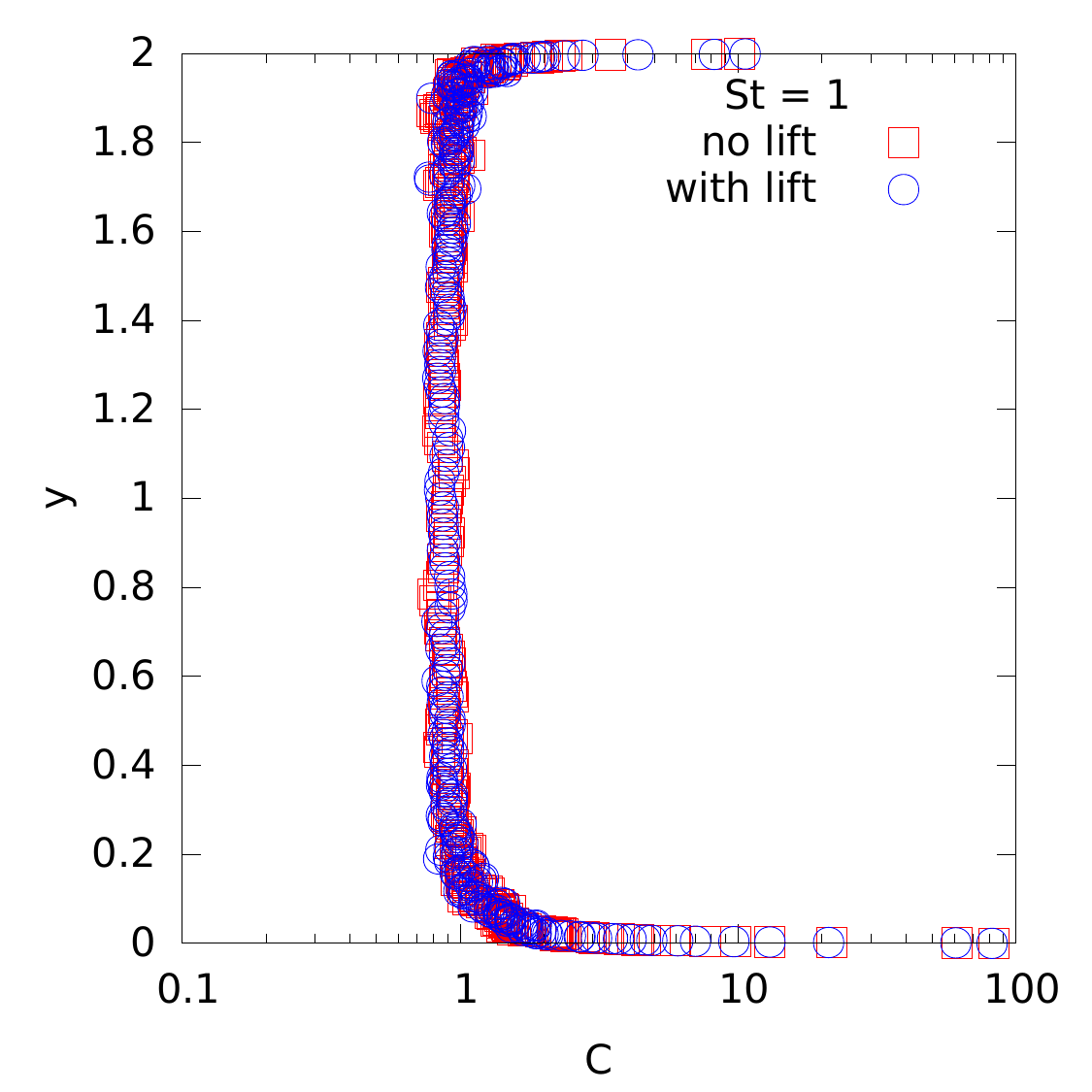}   
{\scriptsize \put(-152,140){\bf (a)}}
\includegraphics[width=0.32\linewidth] {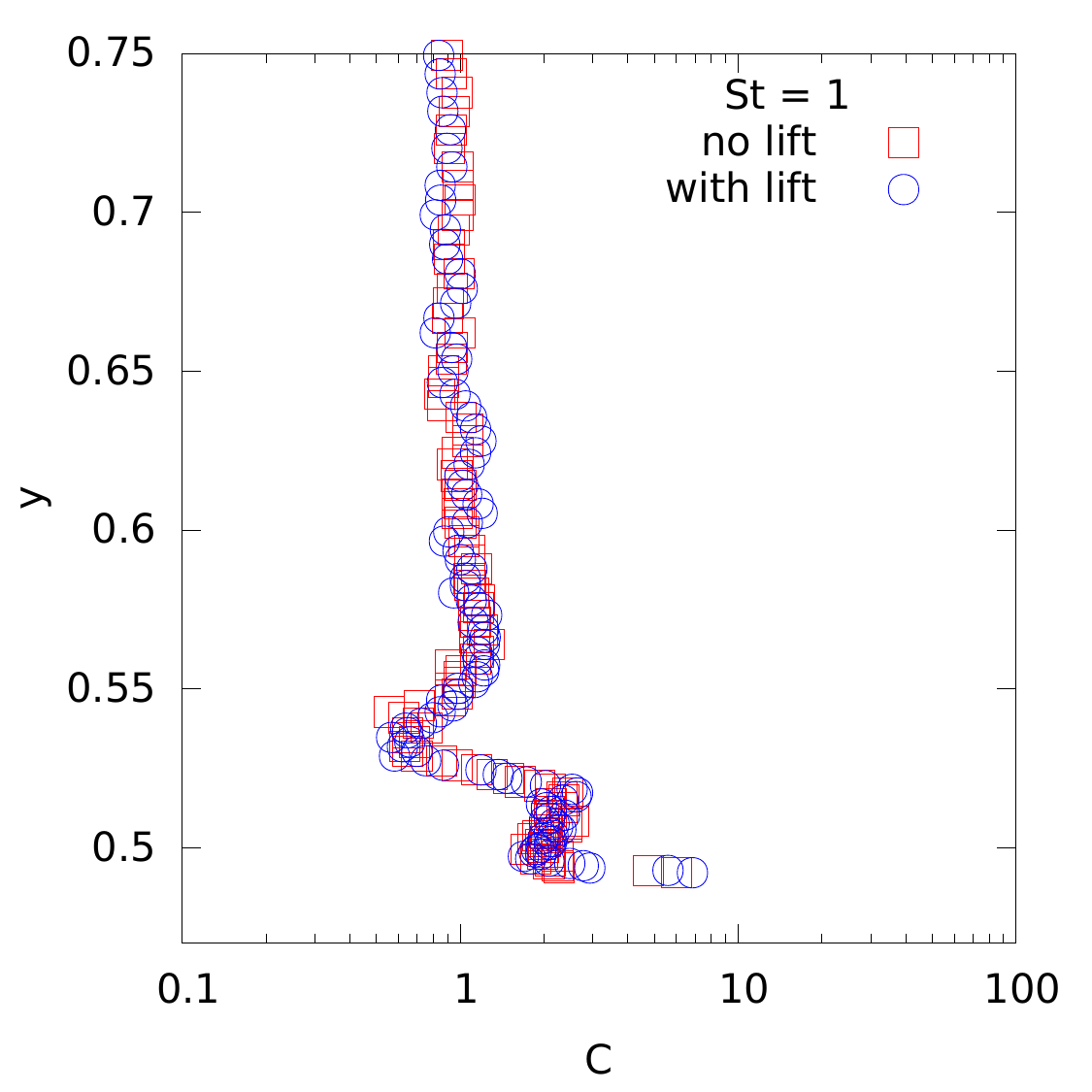}
{\scriptsize \put(-152,140){\bf (b)}}
\includegraphics[width=0.32\linewidth] {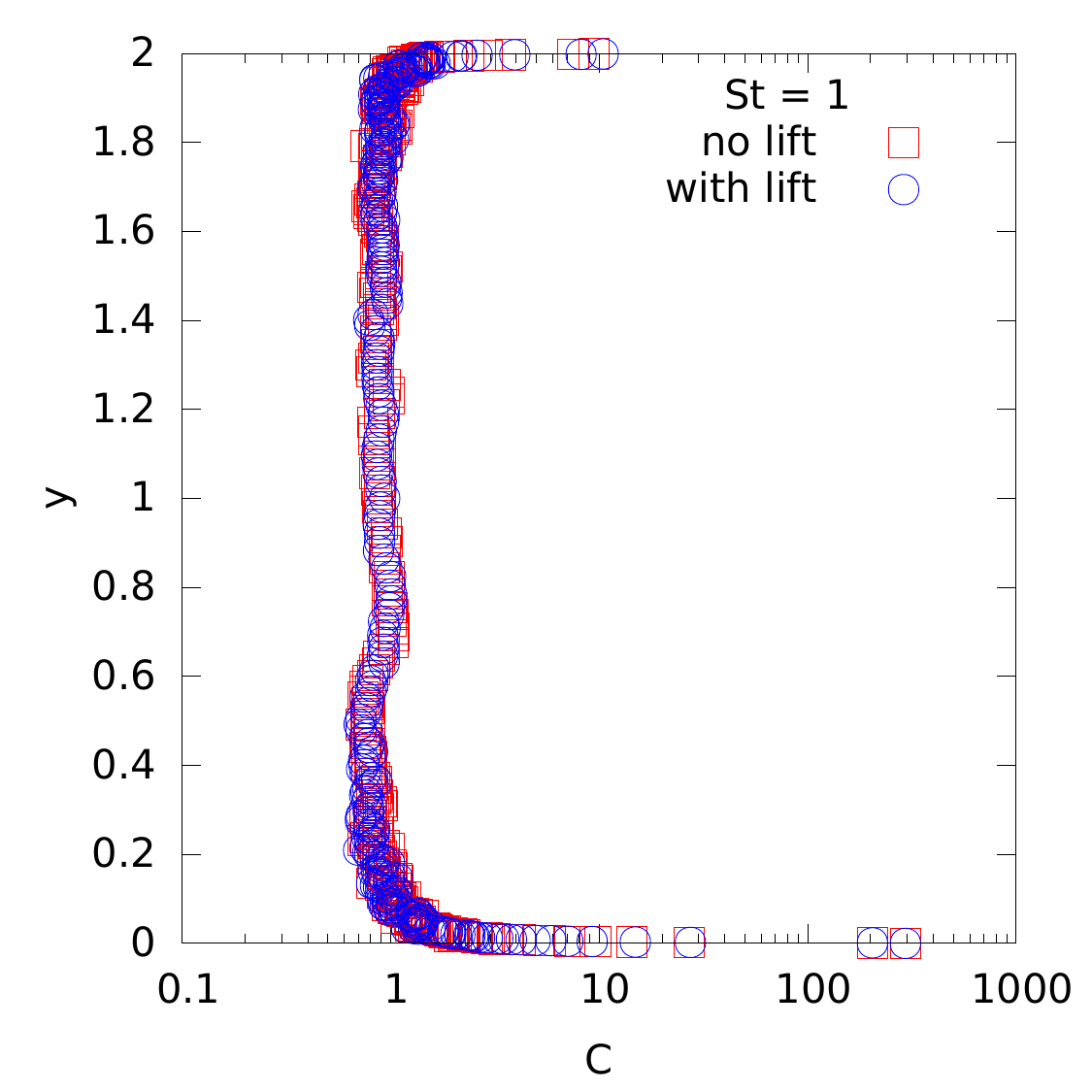}
{\scriptsize \put(-152,140){\bf (c)}}}
\caption{Mean particle concentration for particles having $St^+=1$ at $x=[2.0,3.1,4.0]$ in panels (a) to (c) respectively. 
Panel (b) shows a limited $y$ range to zoom in the area just after the tip of the bump. Blue circles and red squares show 
the particles evolved with and without the lift force respectively.}
\label{fig:conc_st1}
\end{figure}
\begin{figure}
\centering{
\includegraphics[width=0.8\linewidth] {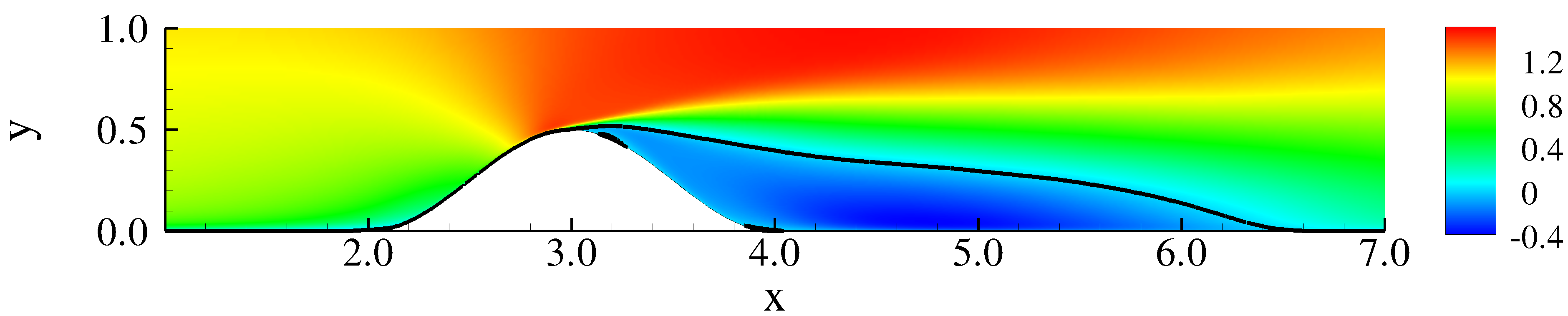}
{\scriptsize \put(-375,67){\bf (a)}}
\includegraphics[width=0.8\linewidth] {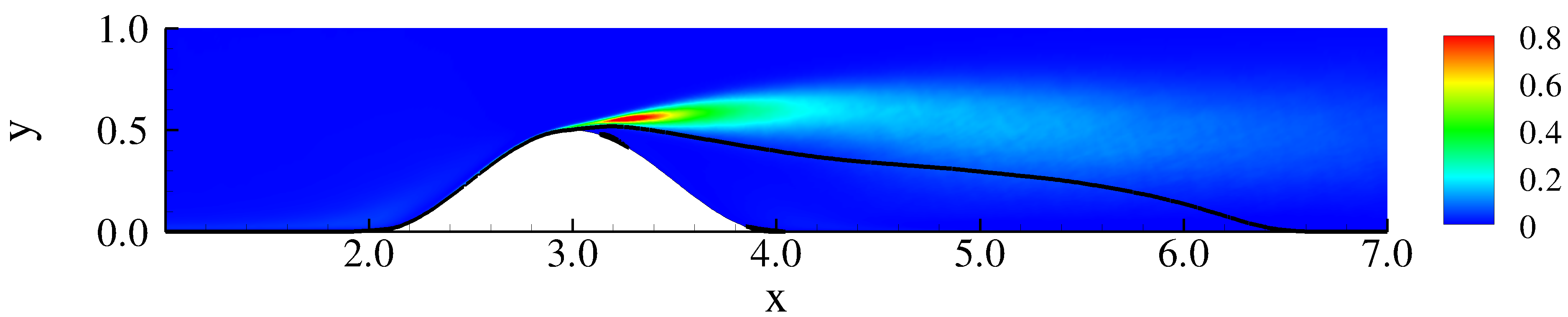}
{\scriptsize \put(-375,67){\bf (b)}} }
\caption{ 
Panel (a): Mean streamwise velocity $\langle u_x \rangle$. 
Panel (b): Mean turbulent kinetic energy production $\Pi$. 
Black solid isoline shows $\langle u_x \rangle = 0$. }
\label{fig:Umean_prod}
\end{figure}

At $St_+=1$, some particles segregate towards the channel walls whilst a homogenous concentration, indicated by the green 
colour, is present in most of the domain. The concentration increases at the bump wall, particularly in three locations: before the 
bump, just after the tip of the bump and after the bump. These locations coincide with three small recirculating 
bubbles that form in the fluid, see the zero velocity isoline in figure~\ref{fig:Umean_prod}(a), that capture the particles. 
Figure~\ref{fig:conc_st1} shows line plots of mean particle concentration at these locations, specifically at $x=[2.0,3.1,4.0]$.
$C=1$ away from the walls for all three $x$ positions, confirming the homogeneous distribution. The concentration increases 
to the order of 10 at the top wall and order 100 at the bottom wall, both before and after the bump, panels (a) and (c). 
In the latter, the particles manage to enter the recirculating region and the concentration only decreases slightly with respect 
to $C=1$ up to $y\approx0.6$. A more pronounced decrease, even though the particles are still clearly present, can be seen 
further behind the bump, but still in the recirculating region, shown by the cyan colour in figure~\ref{fig:conc_zoom}(a).
Just after the tip of the bump, the particles are affected by the shear layer forming in the fluid. 
Figure~\ref{fig:Umean_prod}(b) shows the turbulent kinetic energy production 
${\Pi}= \langle u'_i u'_j\rangle ( {\partial \langle u_i \rangle }/{ \partial x_j} )$, where angular brackets and 
apices indicate the mean and the fluctuations, respectively. The high value of $\Pi$ indicates the location 
and intensity of the shear layer. Figure~\ref{fig:conc_st1} reports the wall normal profiles of the particle concentration at different streamwise position, in particularpanel (b) refers to a streamwise position that traverses the shear layer. The plot shows the accumulation of particles at the bump wall, $y\approx0.49$, up to $y\approx0.53$. 
These are the particles 
that move up the bump wall and towards the tip, since they follow the fluid which is recirculating under the 
shear layer behind the bump, see the negative particle velocity in figure~\ref{fig:3D_vel}(b) and the 
region of negative fluid velocity in figure~\ref{fig:Umean_prod}(a). As the shear layer is encountered, the concentration slightly drops below $C=1$ that is reached when moving further away from the shear layer and towards the center of the channel. For particles having this Stokes number, the effect of the lift is negligible.

%

At $St_+=2$, see figure \ref{fig:conc}(b) and the zoom in figure \ref{fig:conc_zoom}(b), the particle behaviour 
is similar to the one for $St_+=1$, except for the significant decrease in concentration away from the walls 
which is compensated by an increase at the walls. The concentration also decreases in the stream above the 
shear layer with respect to the previous Stokes number. The inertia of the particles increases with Stokes 
number and consequently less particles are capable of entering the recirculating region. The 
increased particle segregation towards the wall persists downstream of the bump where
$C\approx0.5$, and it is more intense than the lower Stokes number case.

\begin{figure}
\centering{
\includegraphics[width=0.32\linewidth] {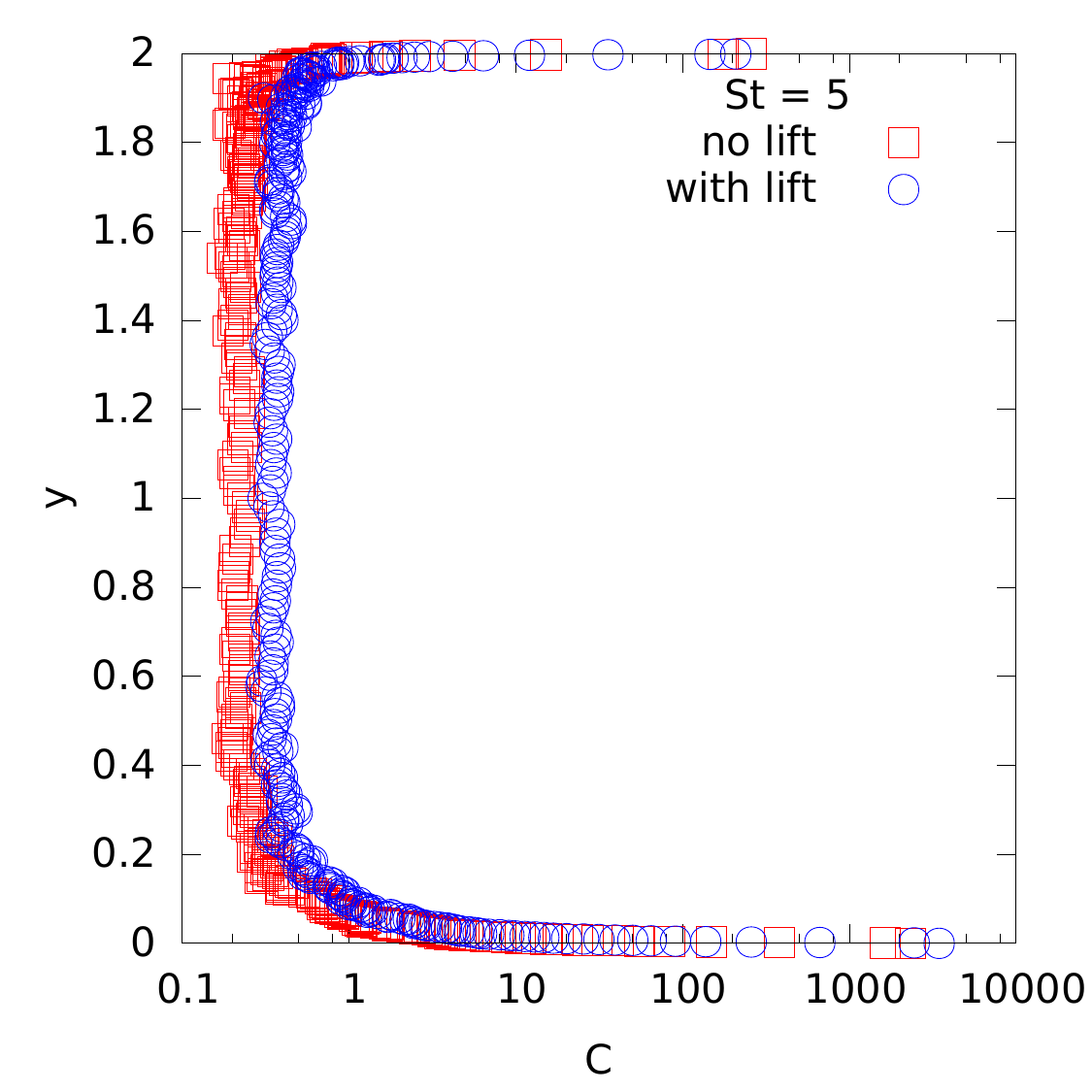}   
{\scriptsize \put(-152,140){\bf (a)}}
\includegraphics[width=0.32\linewidth] {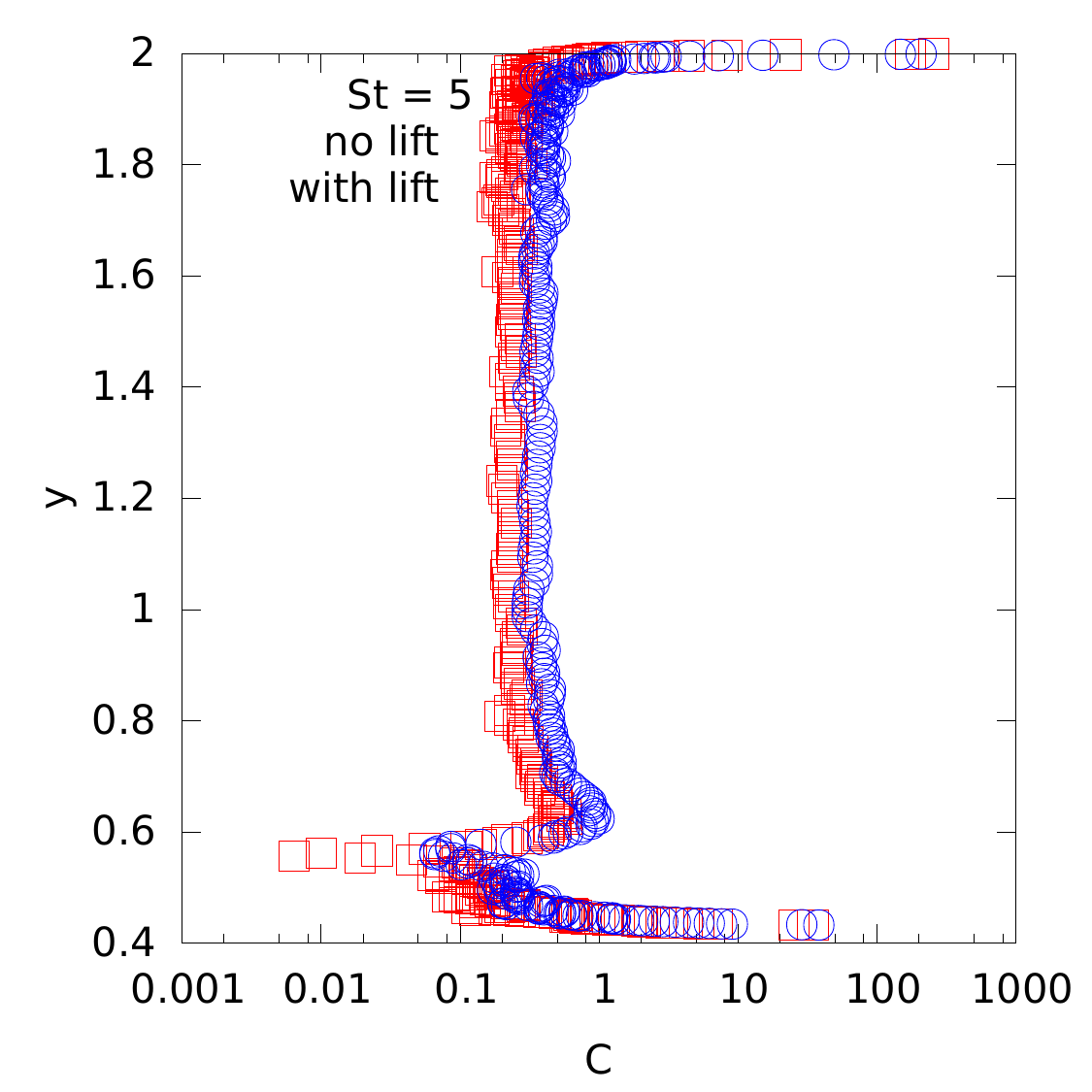}
{\scriptsize \put(-152,140){\bf (b)}}
\includegraphics[width=0.32\linewidth] {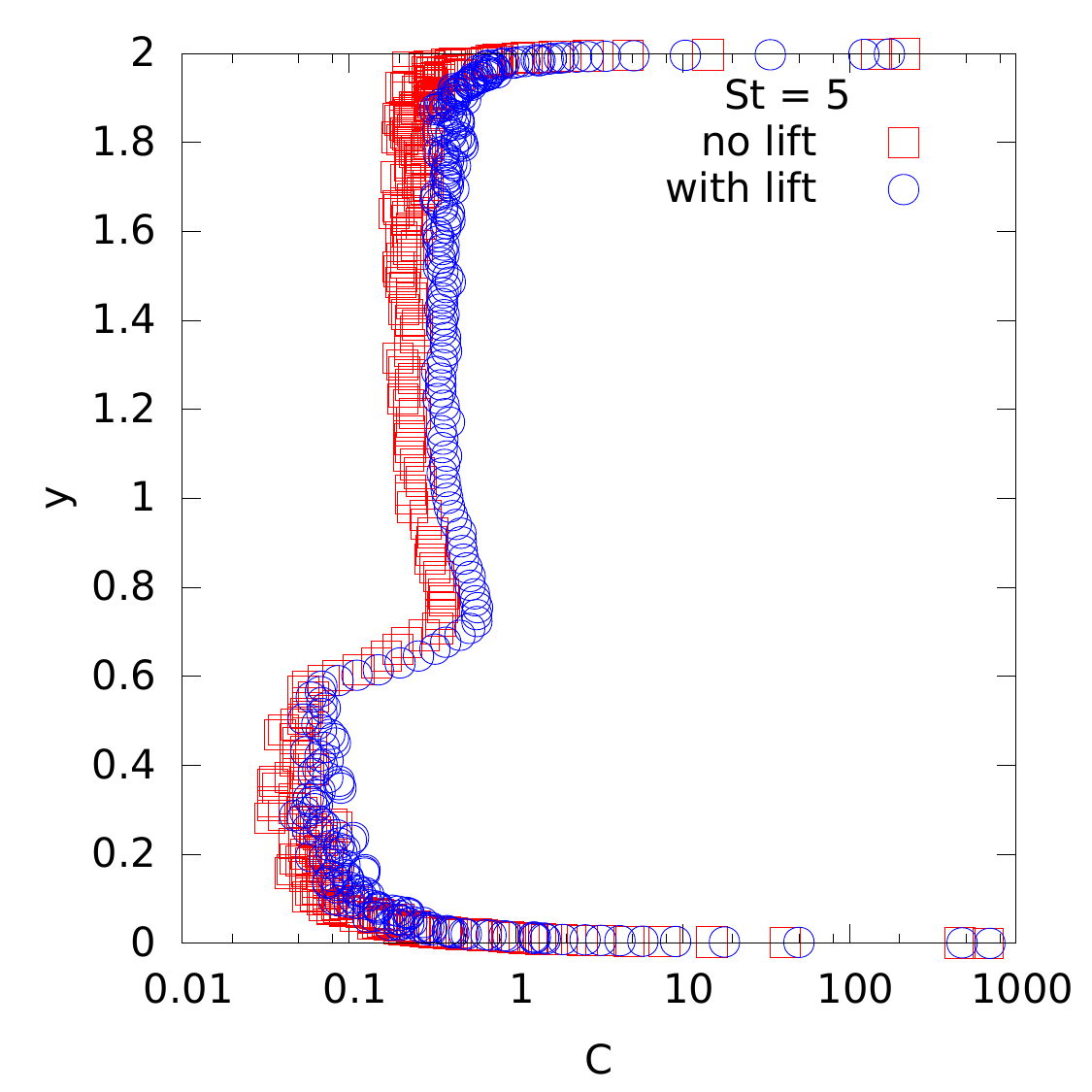}
{\scriptsize \put(-152,140){\bf (c)}}}
\caption{Mean particle concentration for particles having $St^+=5$ at $x=[2.0,3.25,4.0]$ in panels (a) to (c) respectively. 
Blue circles and red squares show the particles evolved with and without the lift force respectively.}
\label{fig:conc_st5}
\end{figure}
When the Stokes number is increased to $St_+=5$, figures \ref{fig:conc}(c) and \ref{fig:conc_zoom}(c), 
the particles segregate more towards the walls and have a low concentration (dark blue) in the rest of the domain. The main recirculating region behind the bump contains no particles (purple colour) except for concentrations of particles in the two, small, secondary recirculating bubbles at the walls (discussed for $St_+=1$). A high concentration of particles, in the order of a thousand times the homogeneous value, is present 
before the bump, at the wall ($x\approx2$). The concentration is high along the bump wall (left side) up to the bump's tip, after which 
the particles are projected towards the centre of the channel by the flow and transported downstream. Figure~\ref{fig:conc_st5} shows 
the mean particle concentration at $x=[2.0,3.25,4.0]$. In general, a slight difference in concentration appears when the lift term is 
included or not, except for an evident deviation from the two values in the shear layer, see panel (b) at $y\approx0.55$. The strong 
vorticity present in the shear layer, which influences the lift, is a direct responsible for this 
discrepancy. Panel (c) shows that the concentration is low in the recirculating region, $y\le0.6$, but nonetheless high at both walls, 
even if the bottom wall is underneath a region containing no particles. The particles must therefore reach this region from the edge 
of the recirculation where the flow reattaches and they are transported towards the left along the wall by reverse (upstream) flow.

At higher Stokes numbers, the particles' inertia becomes dominant and therefore their response to the flow is minimal. 
Figures~\ref{fig:conc}(d), (e) and (f), together with the corresponding panels in figure~\ref{fig:conc_zoom}, show the concentration 
for $St_+=[50, 200, 600]$. After hitting the bump and being projected upwards, the particles proceed towards the upper wall. 
At $St_+=50$, the stream of particles still bends downwards towards the centre of the channel, since the 
mean flow has some effect on them since the characteristic time of the mean flow is comparable with the 
particle relaxation time which in turn is sensibly larger than the fluctuation characteristic time. 
At $St_+=200$, the stream starts resembling a straight line and the rebound of particles at the upper 
wall starts to show. 
Once they rebound, $x\approx3.1$, the flow re-directs them in the streamwise direction and the particles do not enter the 
recirculating region from above. This is not the case for the highest Stokes number, $St_+=600$, since the particles 
are almost entirely inertial and, after bouncing off the top wall, proceed into the recirculating region and create a focused region  
of high concentration at $x\approx4$ close to the bottom wall. The focusing is possible due to the particles hitting the rounded shape 
of the bump on its right-hand side. The particles then spread out back into the channel, depending on the angle at which they hit the 
bump which is a consequence of the angle at which they would have bounced off the top wall from the stream coming 
from the left-hand side of bump. As they initially hit the bump, the particles create a reflection layer, a term coined by \cite{huang_2012}, 
which resembles a shockwave since the population of particles act as a compressible phase. 
The particles segregate towards the upper part of the channel for these higher Stokes numbers. Apart from the particles that  
are projected upwards by the bump, another contribution to this segregation is due to the particles that travel at $y > 0.5$ and 
therefore continue in their trajectory without hitting the bump, with only a minimal effect of the turbulent fluctuations that are not able 
to re-distribute them into all the domain as with the particles at lower Stokes numbers. 

\begin{figure}
\centering
\includegraphics[width=0.49\linewidth] {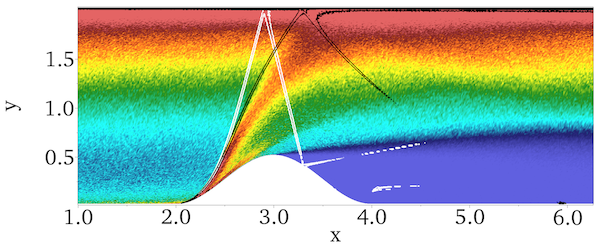}
{\scriptsize \put(-228,87){\bf (a)}}
\includegraphics[width=0.49\linewidth] {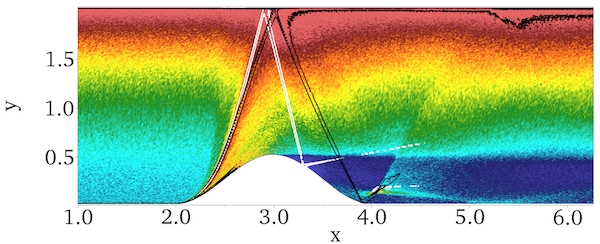}
{\scriptsize \put(-228,87){\bf (b)}}\\
\flushright \includegraphics[width=0.18\linewidth] {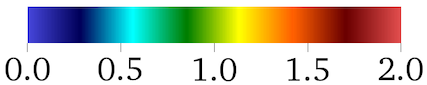}\\
\caption{ 
Mean particle concentration in the fully turbulent flow as coloured contour. The white iso-surface represents the particle 
concentration with no fluid interaction. The black iso-surface represents the particle concentration when interacting  
only with the mean flow. $St_+=200$ in panel (a) and $St_+=600$ in panel (b). Both iso-surfaces are set at $C=7$.}
\label{fig:nofluid_mean}
\end{figure}

Figure~\ref{fig:nofluid_mean} shows to what extent the particle stream is affected by the mean flow or the turbulent flow at these higher 
Stokes numbers. Panels (a) and (b) show $St_+ = 200$ and $St_+ = 600$ respectively with the colour contour representing the mean 
particle concentration for the fully turbulent DNS simulation as in panels (e) and (f) in figure~\ref{fig:conc_zoom}. The white iso-surface 
shows the particle concentration of a separate simulation when there is no coupling, i.e. the particle acceleration is zero and only the rebound from the solid walls is considered. Note that the white 
iso-contours are identical in both panels since the dynamics is independent of the Stokes number (no fluid interaction). The particles are 
given an initial streamwise velocity and the ones travelling at $y<0.5$ hit the bump and create a well-defined reflection layer that bounces 
off the top wall and then the right-hand side of the bump ($x\approx3.3$). Concentration of particles close to the wall at $x\approx4$ is still observed due to the other particles that bounce off the top wall, 
reflect along the back of the bump ($3.5<x<4.0$) and focus in this small region. 
The black iso-surface shows the particle concentration when they are coupled with only the mean flow (no fluctuations) obtained from 
the turbulent simulation. At $St_+ = 200$, the black iso-contour departs from the white one both after the particles are deflected by the bump 
and, most notably, after they bounce off the top wall, disappearing well above the recirculating region. This shows that the mean flow still affects 
the particles by shifting them in the streamwise direction and stops them from entering the recirculation. On the other hand, at $St_+ = 600$, 
the effect of the mean flow is less pronounced. After the deflection by the bump, the three contours are 
almost superimposed.  
After bouncing off the top wall the black iso-contour is slightly deflected but now manages to enter the 
recirculating region. The particles bounce off the bump wall and focus at $x\approx4$,
superimposing the white and coloured contours. 
The turbulent fluctuations therefore play an important role by deviating the particle stream or dispersing it. 
Nonetheless, in high Stokes number cases, considering only the mean flow for particle transport, the 
qualitative behaviour is well predicted.

%

\begin{figure}
\centering{
\includegraphics[width=0.32\linewidth] {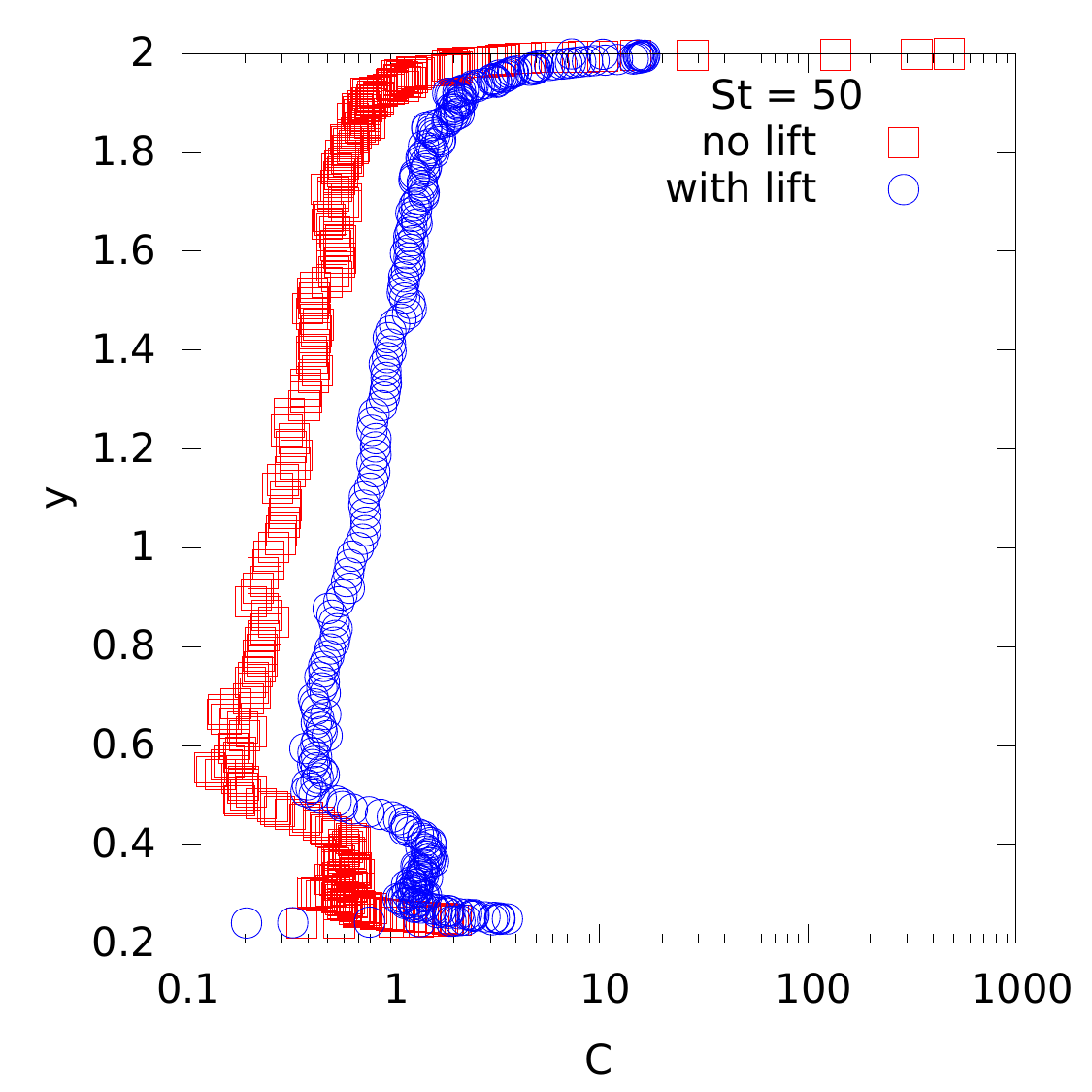}   
{\scriptsize \put(-152,140){\bf (a)}}
\includegraphics[width=0.32\linewidth] {./FIGURES/conc_stk_50_pos5.eps}     
{\scriptsize \put(-152,140){\bf (b)}}
\includegraphics[width=0.32\linewidth] {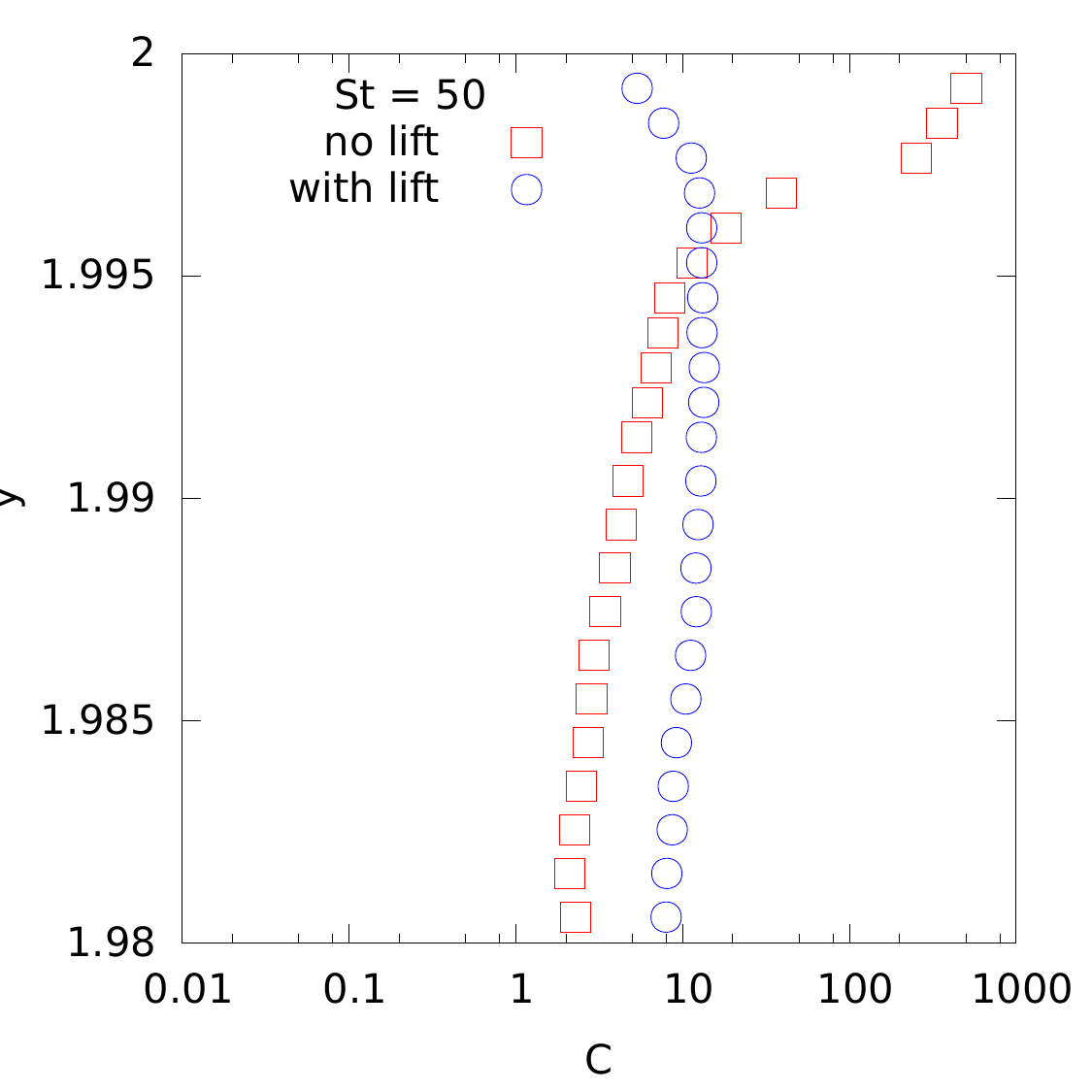}     
{\scriptsize \put(-152,140){\bf (c)}}}
\caption{Mean particle concentration for particles having $St^+=50$ at $x=2.5$ and $x=3.0$ in panels (a) and (b) respectively. 
Panel (c) shows the detail at the top wall at $x=2.5$. Blue circles and red squares show 
the particles evolved with and without the lift force respectively}
\label{fig:conc_st50}
\end{figure}

The lift force plays a crucial role in determining the particle concentration at $St_+=50$, see figure~\ref{fig:conc_st50}. Away from the walls, the concentration for the 
particles without the lift is approximately half the one for particles with the lift and 
therefore the lift  cannot be considered to be negligible, see panels (a) and (b) for plots taken at $x=2.0$ and $x=2.5$ respectively. 
This concentration difference is balanced by a surprising change in concentration at the walls. Panel (c) shows the detail in 
very close proximity of the wall at $x=2.0$ as an example. The concentration for the particles with the lift force is hundreds of 
times lower than those with no lift force. 
Even though this only occurs in a small region close to the wall, the significant quantitative 
difference together with the importance of particle segregation at the wall confirms the importance of the lift force. 
This difference can be again attributed to vorticity which is strong in this region due to the presence of the bounding wall. 
The particles migrating towards the walls experience a negative slip velocity with respect to the fluid, 
$\left.u\right|_p -v_p < 0$ and a negative
fluid vorticity at the bottom wall (positive at the top wall), producing a net lift force towards the center of the 
channel, which prevents the particle wall segregation. 

\begin{figure}
\centering{
\includegraphics[width=0.32\linewidth] {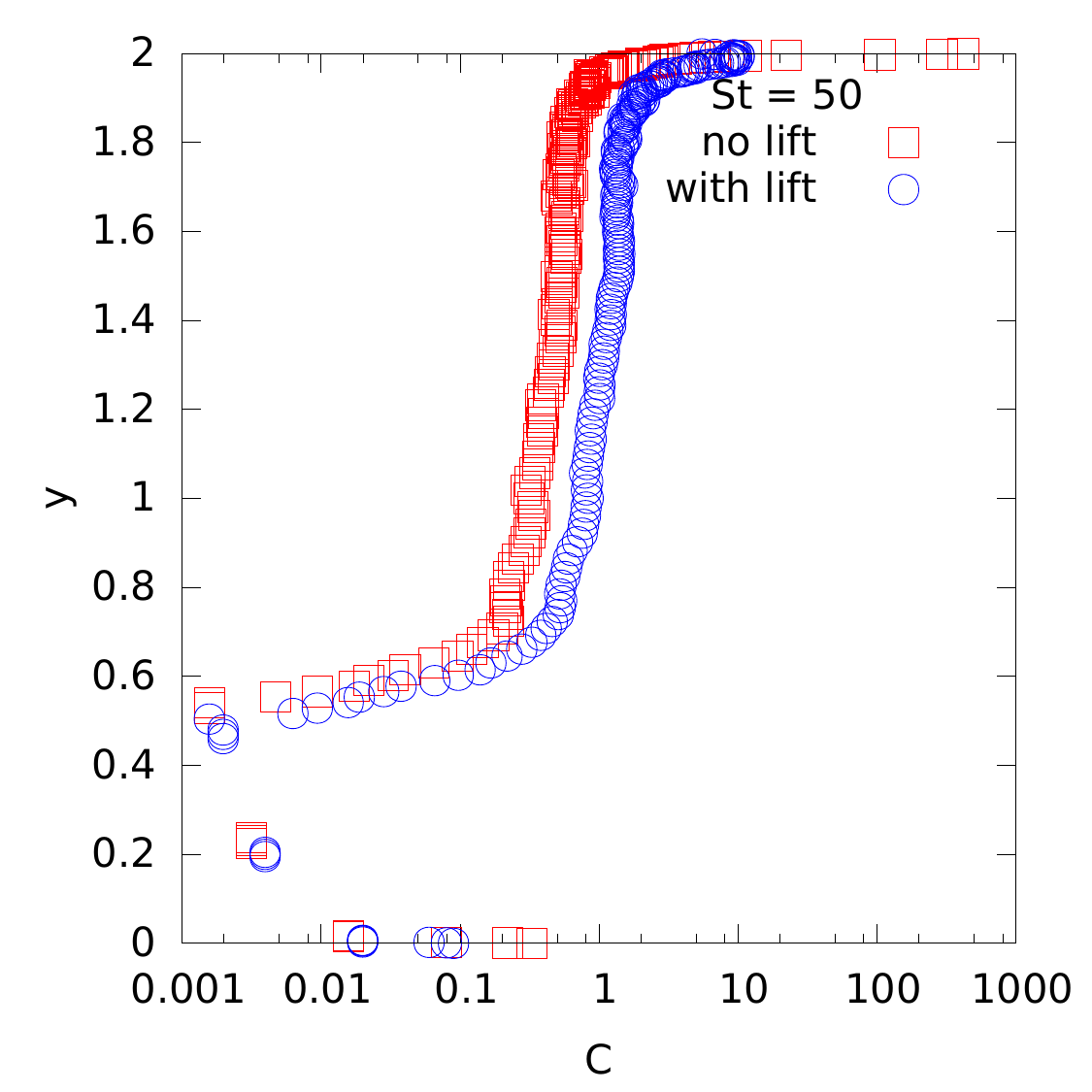} 
{\scriptsize \put(-152,140){\bf (a)}}
\includegraphics[width=0.32\linewidth] {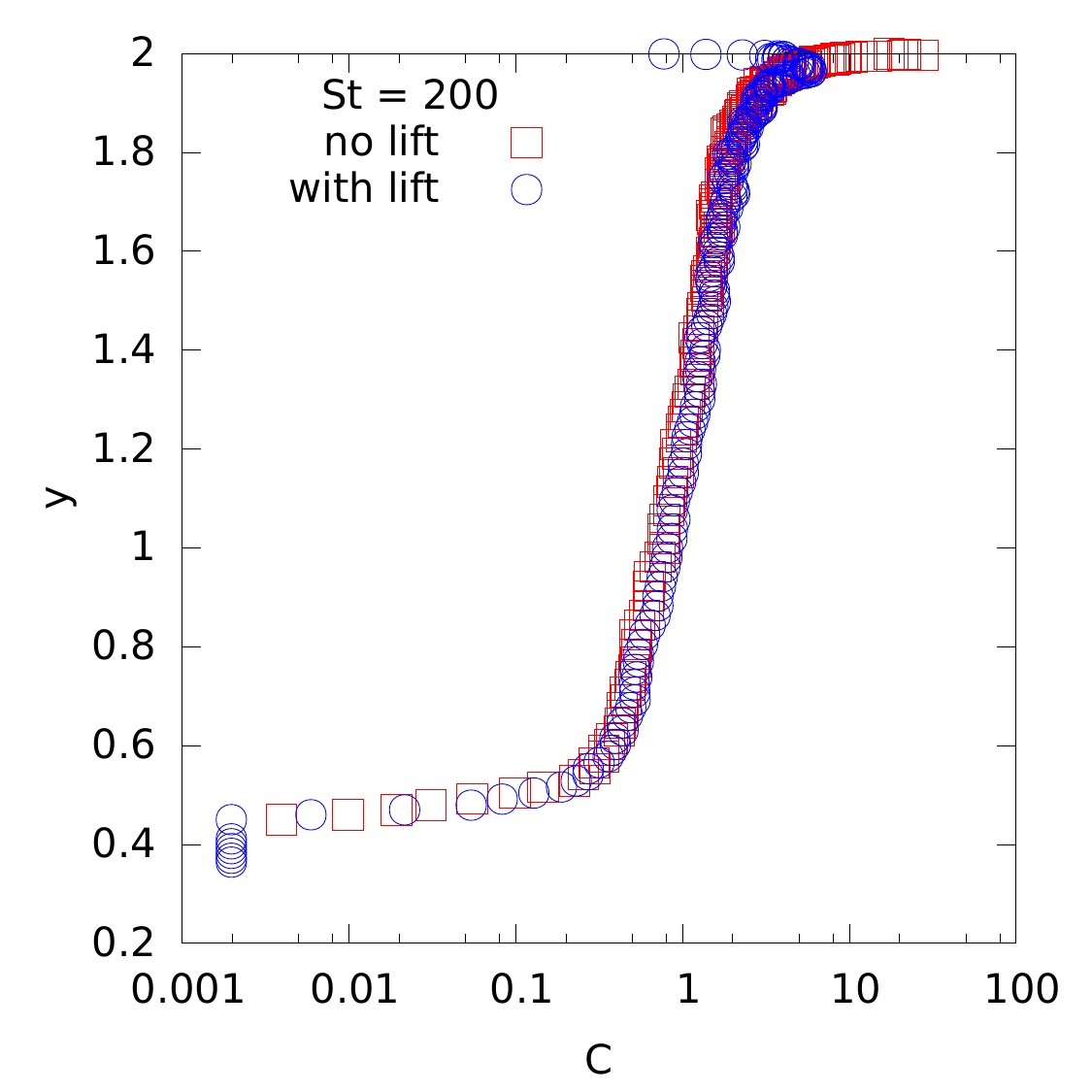}
{\scriptsize \put(-152,140){\bf (b)}}
\includegraphics[width=0.32\linewidth] {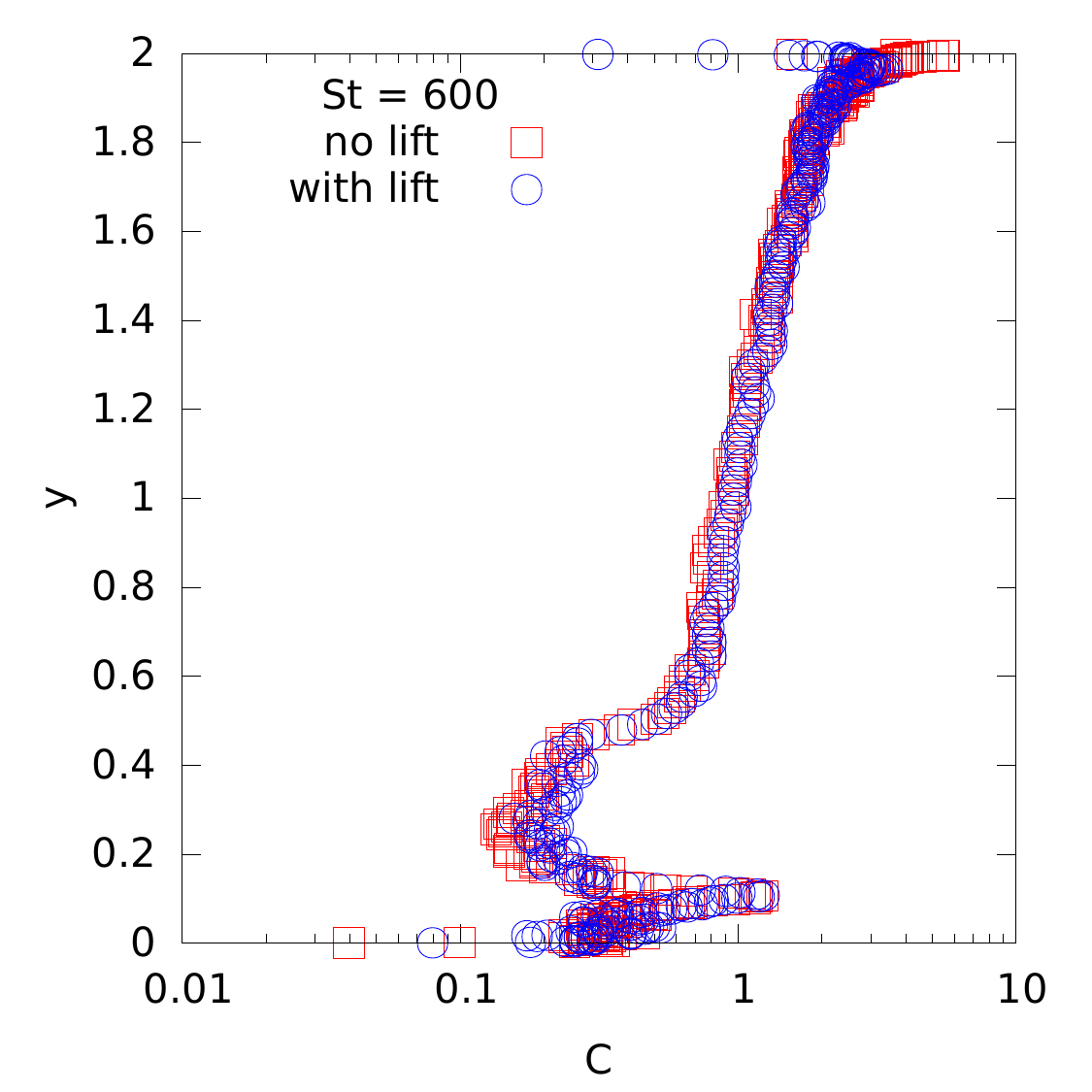}
{\scriptsize \put(-152,140){\bf (c)}}}
\caption{Mean particle concentration at $x=4$ for particles having $St^+=[50,200,600]$ in panels (a) to (c) respectively. 
Blue circles and red squares show the particles evolved with and without the lift force respectively.}
\label{fig:conc_st50_200_600}
\end{figure}

The number of particles at the bottom wall under the recirculating region ($x\approx4$) also decreases significantly compared to the lower Stokes numbers, 
see figure~\ref{fig:conc_st50_200_600}(a). The reverse flow on the right of the recirculating region is not 
able to capture as many particles as previously seen for the lower Stokes numbers since the particles' inertia is now higher. 
When the Stokes increases to $St_+=200$, see panel (b), no particles are captured and the concentration at $y<0.5$ is practically zero.
On the other hand, at $St_+ = 600$, see panel (c), there is a concentration peak inside recirculating region due to the particles bouncing 
off the upper wall and focusing in this region, as discussed previously. The difference in particle concentration with or without the lift force 
disappears away from the walls as the Stokes number is increased but still persists at the top wall where particles with lift force have 
significantly less concentration.

%
%


\section{Final remarks}
\label{sec:final}

The effects of Stokes number and lift force have been analysed for particle-laden turbulent flow that separates 
due to the presence of a bump in a channel-like domain. A strong shear layer and a recirculating region are 
formed behind the bump, both affecting the particle dynamics. The Reynolds number is relatively high when 
considering similar geometries and the present multi-phase configuration has, to the best of our knowledge, 
never been simulated before.

A vast range of Stokes numbers are considered, simulated both with and without the lift force in the particle 
dynamic equation. The conclusion is that the lift force must not be neglected, since 
there are drastic changes in particle concentration for some Stokes numbers with respect to results obtained 
without the lift force. Regions of high vorticity, particularly at the walls and in the shear layer, exhibit the greatest 
differences. However, for some intermediate Stokes numbers, the difference is also evident in the bulk of the 
flow throughout the whole domain.

The particles behave as tracers at low Stokes numbers and closely follow the fluid phase. As the Stokes number 
is increased, the particles tend to segregate at the walls and do not enter the recirculating region behind the bump. 
Some particles are captured by the recirculation, forced upstream as they move close to the wall and transported 
back downstream as they encounter the strong shear layer formed by the bump. Secondary recirculating regions, 
one before the bump and two inside the primary recirculating region, manage to capture the particles. At the highest 
Stokes numbers, the particles' inertia is high and their ballistic nature makes them bounce off the bump, top and 
bottom walls, creating reflection layers as they are only slightly affected by the fluid flow. This enables the particles 
to enter  the recirculating region by bouncing off the back of the bump and creating a focused spot of high particle 
concentration.

Understanding particle-laden flows in presence of features such as a bump is important for engineering applications 
that concern geometries relatively more complex than classical flows such as straight pipes or planar channels. Such 
features may be intentional (such as for filtration or separation of particles) or unintentional (such as defects) and it is 
essential to comprehend how particles, that may have different Stokes numbers, behave in such particle-laden flows.




\begin{acknowledgments}
The research has received funding from the European Research Council under the ERC Grant 
Agreement no. 339446. We acknowledge CINECA for awarding us access to supercomputing 
resource MARCONI based in Bologna, Italy through ISCRA project no. HP10BLVPKA. 
\end{acknowledgments}

\bibliographystyle{aipnum4-1}
\bibliography{bump_particles_pof}

\providecommand{\noopsort}[1]{}\providecommand{\singleletter}[1]{#1}%
\begin{thebibliography}{56}%
\makeatletter
\providecommand \@ifxundefined [1]{%
 \@ifx{#1\undefined}
}%
\providecommand \@ifnum [1]{%
 \ifnum #1\expandafter \@firstoftwo
 \else \expandafter \@secondoftwo
 \fi
}%
\providecommand \@ifx [1]{%
 \ifx #1\expandafter \@firstoftwo
 \else \expandafter \@secondoftwo
 \fi
}%
\providecommand \natexlab [1]{#1}%
\providecommand \enquote  [1]{``#1''}%
\providecommand \bibnamefont  [1]{#1}%
\providecommand \bibfnamefont [1]{#1}%
\providecommand \citenamefont [1]{#1}%
\providecommand \href@noop [0]{\@secondoftwo}%
\providecommand \href [0]{\begingroup \@sanitize@url \@href}%
\providecommand \@href[1]{\@@startlink{#1}\@@href}%
\providecommand \@@href[1]{\endgroup#1\@@endlink}%
\providecommand \@sanitize@url [0]{\catcode `\\12\catcode `\$12\catcode
  `\&12\catcode `\#12\catcode `\^12\catcode `\_12\catcode `\%12\relax}%
\providecommand \@@startlink[1]{}%
\providecommand \@@endlink[0]{}%
\providecommand \url  [0]{\begingroup\@sanitize@url \@url }%
\providecommand \@url [1]{\endgroup\@href {#1}{\urlprefix }}%
\providecommand \urlprefix  [0]{URL }%
\providecommand \Eprint [0]{\href }%
\providecommand \doibase [0]{http://dx.doi.org/}%
\providecommand \selectlanguage [0]{\@gobble}%
\providecommand \bibinfo  [0]{\@secondoftwo}%
\providecommand \bibfield  [0]{\@secondoftwo}%
\providecommand \translation [1]{[#1]}%
\providecommand \BibitemOpen [0]{}%
\providecommand \bibitemStop [0]{}%
\providecommand \bibitemNoStop [0]{.\EOS\space}%
\providecommand \EOS [0]{\spacefactor3000\relax}%
\providecommand \BibitemShut  [1]{\csname bibitem#1\endcsname}%
\let\auto@bib@innerbib\@empty
\bibitem [{\citenamefont {Balachandar}\ and\ \citenamefont
  {Eaton}(2010)}]{balachandar_2010}%
  \BibitemOpen
  \bibfield  {author} {\bibinfo {author} {\bibfnamefont {S.}~\bibnamefont
  {Balachandar}}\ and\ \bibinfo {author} {\bibfnamefont {J.~K.}\ \bibnamefont
  {Eaton}},\ }\bibfield  {title} {\enquote {\bibinfo {title} {Turbulent
  dispersed multiphase flow},}\ }\href@noop {} {\bibfield  {journal} {\bibinfo
  {journal} {Annual Review of Fluid Mechanics}\ }\textbf {\bibinfo {volume}
  {42}},\ \bibinfo {pages} {111--133} (\bibinfo {year} {2010})}\BibitemShut
  {NoStop}%
\bibitem [{\citenamefont {Elghobashi}(2019)}]{elghobashi2019direct}%
  \BibitemOpen
  \bibfield  {author} {\bibinfo {author} {\bibfnamefont {S.}~\bibnamefont
  {Elghobashi}},\ }\bibfield  {title} {\enquote {\bibinfo {title} {Direct
  numerical simulation of turbulent flows laden with droplets or bubbles},}\
  }\href@noop {} {\bibfield  {journal} {\bibinfo  {journal} {Annual Review of
  Fluid Mechanics}\ }\textbf {\bibinfo {volume} {51}},\ \bibinfo {pages}
  {217--244} (\bibinfo {year} {2019})}\BibitemShut {NoStop}%
\bibitem [{\citenamefont {Marchioli}(2017)}]{marchioli2017large}%
  \BibitemOpen
  \bibfield  {author} {\bibinfo {author} {\bibfnamefont {C.}~\bibnamefont
  {Marchioli}},\ }\bibfield  {title} {\enquote {\bibinfo {title} {Large-eddy
  simulation of turbulent dispersed flows: a review of modelling approaches},}\
  }\href@noop {} {\bibfield  {journal} {\bibinfo  {journal} {Acta Mechanica}\
  }\textbf {\bibinfo {volume} {228}},\ \bibinfo {pages} {741--771} (\bibinfo
  {year} {2017})}\BibitemShut {NoStop}%
\bibitem [{\citenamefont {Innocenti}, \citenamefont {Marchioli},\ and\
  \citenamefont {Chibbaro}(2016)}]{innocenti2016lagrangian}%
  \BibitemOpen
  \bibfield  {author} {\bibinfo {author} {\bibfnamefont {A.}~\bibnamefont
  {Innocenti}}, \bibinfo {author} {\bibfnamefont {C.}~\bibnamefont
  {Marchioli}}, \ and\ \bibinfo {author} {\bibfnamefont {S.}~\bibnamefont
  {Chibbaro}},\ }\bibfield  {title} {\enquote {\bibinfo {title} {Lagrangian
  filtered density function for les-based stochastic modelling of turbulent
  particle-laden flows},}\ }\href@noop {} {\bibfield  {journal} {\bibinfo
  {journal} {Physics of Fluids}\ }\textbf {\bibinfo {volume} {28}},\ \bibinfo
  {pages} {115106} (\bibinfo {year} {2016})}\BibitemShut {NoStop}%
\bibitem [{\citenamefont {Park}\ \emph {et~al.}(2017)\citenamefont {Park},
  \citenamefont {Bassenne}, \citenamefont {Urzay},\ and\ \citenamefont
  {Moin}}]{park2017simple}%
  \BibitemOpen
  \bibfield  {author} {\bibinfo {author} {\bibfnamefont {G.~I.}\ \bibnamefont
  {Park}}, \bibinfo {author} {\bibfnamefont {M.}~\bibnamefont {Bassenne}},
  \bibinfo {author} {\bibfnamefont {J.}~\bibnamefont {Urzay}}, \ and\ \bibinfo
  {author} {\bibfnamefont {P.}~\bibnamefont {Moin}},\ }\bibfield  {title}
  {\enquote {\bibinfo {title} {A simple dynamic subgrid-scale model for les of
  particle-laden turbulence},}\ }\href@noop {} {\bibfield  {journal} {\bibinfo
  {journal} {Physical Review Fluids}\ }\textbf {\bibinfo {volume} {2}},\
  \bibinfo {pages} {044301} (\bibinfo {year} {2017})}\BibitemShut {NoStop}%
\bibitem [{\citenamefont {Minier}, \citenamefont {Chibbaro},\ and\
  \citenamefont {Pope}(2014)}]{minier2014guidelines}%
  \BibitemOpen
  \bibfield  {author} {\bibinfo {author} {\bibfnamefont {J.-P.}\ \bibnamefont
  {Minier}}, \bibinfo {author} {\bibfnamefont {S.}~\bibnamefont {Chibbaro}}, \
  and\ \bibinfo {author} {\bibfnamefont {S.~B.}\ \bibnamefont {Pope}},\
  }\bibfield  {title} {\enquote {\bibinfo {title} {Guidelines for the
  formulation of lagrangian stochastic models for particle simulations of
  single-phase and dispersed two-phase turbulent flows},}\ }\href@noop {}
  {\bibfield  {journal} {\bibinfo  {journal} {Physics of Fluids}\ }\textbf
  {\bibinfo {volume} {26}},\ \bibinfo {pages} {113303} (\bibinfo {year}
  {2014})}\BibitemShut {NoStop}%
\bibitem [{\citenamefont {Sajjadi}\ \emph {et~al.}(2017)\citenamefont
  {Sajjadi}, \citenamefont {Salmanzadeh}, \citenamefont {Ahmadi},\ and\
  \citenamefont {Jafari}}]{sajjadi2017lattice}%
  \BibitemOpen
  \bibfield  {author} {\bibinfo {author} {\bibfnamefont {H.}~\bibnamefont
  {Sajjadi}}, \bibinfo {author} {\bibfnamefont {M.}~\bibnamefont
  {Salmanzadeh}}, \bibinfo {author} {\bibfnamefont {G.}~\bibnamefont {Ahmadi}},
  \ and\ \bibinfo {author} {\bibfnamefont {S.}~\bibnamefont {Jafari}},\
  }\bibfield  {title} {\enquote {\bibinfo {title} {Lattice boltzmann method and
  rans approach for simulation of turbulent flows and particle transport and
  deposition},}\ }\href@noop {} {\bibfield  {journal} {\bibinfo  {journal}
  {Particuology}\ }\textbf {\bibinfo {volume} {30}},\ \bibinfo {pages} {62--72}
  (\bibinfo {year} {2017})}\BibitemShut {NoStop}%
\bibitem [{\citenamefont {Vahidifar}, \citenamefont {Saffarian},\ and\
  \citenamefont {Hajidavalloo}(2018)}]{Vahidifar2018}%
  \BibitemOpen
  \bibfield  {author} {\bibinfo {author} {\bibfnamefont {S.}~\bibnamefont
  {Vahidifar}}, \bibinfo {author} {\bibfnamefont {M.~R.}\ \bibnamefont
  {Saffarian}}, \ and\ \bibinfo {author} {\bibfnamefont {E.}~\bibnamefont
  {Hajidavalloo}},\ }\bibfield  {title} {\enquote {\bibinfo {title}
  {Introducing the theory of successful settling in order to evaluate and
  optimize the sedimentation tanks},}\ }\href {\doibase
  10.1007/s11012-018-0907-2} {\bibfield  {journal} {\bibinfo  {journal}
  {Meccanica}\ } (\bibinfo {year} {2018}),\
  10.1007/s11012-018-0907-2}\BibitemShut {NoStop}%
\bibitem [{\citenamefont {Elghobashi}(1994)}]{elghobashi1994predicting}%
  \BibitemOpen
  \bibfield  {author} {\bibinfo {author} {\bibfnamefont {S.}~\bibnamefont
  {Elghobashi}},\ }\bibfield  {title} {\enquote {\bibinfo {title} {On
  predicting particle-laden turbulent flows},}\ }\href@noop {} {\bibfield
  {journal} {\bibinfo  {journal} {Applied Scientific Research}\ }\textbf
  {\bibinfo {volume} {52}},\ \bibinfo {pages} {309--329} (\bibinfo {year}
  {1994})}\BibitemShut {NoStop}%
\bibitem [{\citenamefont {Toschi}\ and\ \citenamefont
  {Bodenschatz}(2009)}]{toschi_2009}%
  \BibitemOpen
  \bibfield  {author} {\bibinfo {author} {\bibfnamefont {F.}~\bibnamefont
  {Toschi}}\ and\ \bibinfo {author} {\bibfnamefont {E.}~\bibnamefont
  {Bodenschatz}},\ }\bibfield  {title} {\enquote {\bibinfo {title} {Lagrangian
  properties of particles in turbulence},}\ }\href@noop {} {\bibfield
  {journal} {\bibinfo  {journal} {Annual review of fluid mechanics}\ }\textbf
  {\bibinfo {volume} {41}},\ \bibinfo {pages} {375--404} (\bibinfo {year}
  {2009})}\BibitemShut {NoStop}%
\bibitem [{\citenamefont {Kuerten}(2016)}]{kuerten2016point}%
  \BibitemOpen
  \bibfield  {author} {\bibinfo {author} {\bibfnamefont {J.~G.~M.}\
  \bibnamefont {Kuerten}},\ }\bibfield  {title} {\enquote {\bibinfo {title}
  {Point-particle dns and les of particle-laden turbulent flow-a
  state-of-the-art review},}\ }\href@noop {} {\bibfield  {journal} {\bibinfo
  {journal} {Flow, turbulence and combustion}\ }\textbf {\bibinfo {volume}
  {97}},\ \bibinfo {pages} {689--713} (\bibinfo {year} {2016})}\BibitemShut
  {NoStop}%
\bibitem [{\citenamefont {Gustavsson}\ \emph {et~al.}(2017)\citenamefont
  {Gustavsson}, \citenamefont {Jucha}, \citenamefont {Naso}, \citenamefont
  {L{\'e}v{\^e}que}, \citenamefont {Pumir},\ and\ \citenamefont
  {Mehlig}}]{gustavsson2017statistical}%
  \BibitemOpen
  \bibfield  {author} {\bibinfo {author} {\bibfnamefont {K.}~\bibnamefont
  {Gustavsson}}, \bibinfo {author} {\bibfnamefont {J.}~\bibnamefont {Jucha}},
  \bibinfo {author} {\bibfnamefont {A.}~\bibnamefont {Naso}}, \bibinfo {author}
  {\bibfnamefont {E.}~\bibnamefont {L{\'e}v{\^e}que}}, \bibinfo {author}
  {\bibfnamefont {A.}~\bibnamefont {Pumir}}, \ and\ \bibinfo {author}
  {\bibfnamefont {B.}~\bibnamefont {Mehlig}},\ }\bibfield  {title} {\enquote
  {\bibinfo {title} {Statistical model for the orientation of nonspherical
  particles settling in turbulence},}\ }\href@noop {} {\bibfield  {journal}
  {\bibinfo  {journal} {Physical review letters}\ }\textbf {\bibinfo {volume}
  {119}},\ \bibinfo {pages} {254501} (\bibinfo {year} {2017})}\BibitemShut
  {NoStop}%
\bibitem [{\citenamefont {Voth}\ and\ \citenamefont
  {Soldati}(2017)}]{voth2017anisotropic}%
  \BibitemOpen
  \bibfield  {author} {\bibinfo {author} {\bibfnamefont {G.~A.}\ \bibnamefont
  {Voth}}\ and\ \bibinfo {author} {\bibfnamefont {A.}~\bibnamefont {Soldati}},\
  }\bibfield  {title} {\enquote {\bibinfo {title} {Anisotropic particles in
  turbulence},}\ }\href@noop {} {\bibfield  {journal} {\bibinfo  {journal}
  {Annual Review of Fluid Mechanics}\ }\textbf {\bibinfo {volume} {49}},\
  \bibinfo {pages} {249--276} (\bibinfo {year} {2017})}\BibitemShut {NoStop}%
\bibitem [{\citenamefont {Squires}\ and\ \citenamefont
  {Eaton}(1991)}]{squires1991preferential}%
  \BibitemOpen
  \bibfield  {author} {\bibinfo {author} {\bibfnamefont {K.~D.}\ \bibnamefont
  {Squires}}\ and\ \bibinfo {author} {\bibfnamefont {J.~K.}\ \bibnamefont
  {Eaton}},\ }\bibfield  {title} {\enquote {\bibinfo {title} {Preferential
  concentration of particles by turbulence},}\ }\href@noop {} {\bibfield
  {journal} {\bibinfo  {journal} {Physics of Fluids A: Fluid Dynamics}\
  }\textbf {\bibinfo {volume} {3}},\ \bibinfo {pages} {1169--1178} (\bibinfo
  {year} {1991})}\BibitemShut {NoStop}%
\bibitem [{\citenamefont {Bragg}, \citenamefont {Ireland},\ and\ \citenamefont
  {Collins}(2015)}]{bragg2015mechanisms}%
  \BibitemOpen
  \bibfield  {author} {\bibinfo {author} {\bibfnamefont {A.~D.}\ \bibnamefont
  {Bragg}}, \bibinfo {author} {\bibfnamefont {P.~J.}\ \bibnamefont {Ireland}},
  \ and\ \bibinfo {author} {\bibfnamefont {L.~R.}\ \bibnamefont {Collins}},\
  }\bibfield  {title} {\enquote {\bibinfo {title} {Mechanisms for the
  clustering of inertial particles in the inertial range of isotropic
  turbulence},}\ }\href@noop {} {\bibfield  {journal} {\bibinfo  {journal}
  {Physical Review E}\ }\textbf {\bibinfo {volume} {92}},\ \bibinfo {pages}
  {023029} (\bibinfo {year} {2015})}\BibitemShut {NoStop}%
\bibitem [{\citenamefont {Nicolai}\ \emph {et~al.}(2014)\citenamefont
  {Nicolai}, \citenamefont {Jacob}, \citenamefont {Gualtieri},\ and\
  \citenamefont {Piva}}]{nicolai2014inertial}%
  \BibitemOpen
  \bibfield  {author} {\bibinfo {author} {\bibfnamefont {C.}~\bibnamefont
  {Nicolai}}, \bibinfo {author} {\bibfnamefont {B.}~\bibnamefont {Jacob}},
  \bibinfo {author} {\bibfnamefont {P.}~\bibnamefont {Gualtieri}}, \ and\
  \bibinfo {author} {\bibfnamefont {R.}~\bibnamefont {Piva}},\ }\bibfield
  {title} {\enquote {\bibinfo {title} {Inertial particles in homogeneous shear
  turbulence: experiments and direct numerical simulation},}\ }\href@noop {}
  {\bibfield  {journal} {\bibinfo  {journal} {Flow, turbulence and combustion}\
  }\textbf {\bibinfo {volume} {92}},\ \bibinfo {pages} {65--82} (\bibinfo
  {year} {2014})}\BibitemShut {NoStop}%
\bibitem [{\citenamefont {Battista}\ \emph {et~al.}(2018)\citenamefont
  {Battista}, \citenamefont {Gualtieri}, \citenamefont {Mollicone},\ and\
  \citenamefont {Casciola}}]{battista2018application}%
  \BibitemOpen
  \bibfield  {author} {\bibinfo {author} {\bibfnamefont {F.}~\bibnamefont
  {Battista}}, \bibinfo {author} {\bibfnamefont {P.}~\bibnamefont {Gualtieri}},
  \bibinfo {author} {\bibfnamefont {J.-P.}\ \bibnamefont {Mollicone}}, \ and\
  \bibinfo {author} {\bibfnamefont {C.~M.}\ \bibnamefont {Casciola}},\
  }\bibfield  {title} {\enquote {\bibinfo {title} {Application of the exact
  regularized point particle method (erpp) to particle laden turbulent shear
  flows in the two-way coupling regime},}\ }\href@noop {} {\bibfield  {journal}
  {\bibinfo  {journal} {International Journal of Multiphase Flow}\ }\textbf
  {\bibinfo {volume} {101}},\ \bibinfo {pages} {113--124} (\bibinfo {year}
  {2018})}\BibitemShut {NoStop}%
\bibitem [{\citenamefont {Sardina}\ \emph {et~al.}(2011)\citenamefont
  {Sardina}, \citenamefont {Picano}, \citenamefont {Schlatter}, \citenamefont
  {Brandt},\ and\ \citenamefont {Casciola}}]{sardina2011large}%
  \BibitemOpen
  \bibfield  {author} {\bibinfo {author} {\bibfnamefont {G.}~\bibnamefont
  {Sardina}}, \bibinfo {author} {\bibfnamefont {F.}~\bibnamefont {Picano}},
  \bibinfo {author} {\bibfnamefont {P.}~\bibnamefont {Schlatter}}, \bibinfo
  {author} {\bibfnamefont {L.}~\bibnamefont {Brandt}}, \ and\ \bibinfo {author}
  {\bibfnamefont {C.~M.}\ \bibnamefont {Casciola}},\ }\bibfield  {title}
  {\enquote {\bibinfo {title} {Large scale accumulation patterns of inertial
  particles in wall-bounded turbulent flow},}\ }\href@noop {} {\bibfield
  {journal} {\bibinfo  {journal} {Flow, turbulence and combustion}\ }\textbf
  {\bibinfo {volume} {86}},\ \bibinfo {pages} {519--532} (\bibinfo {year}
  {2011})}\BibitemShut {NoStop}%
\bibitem [{\citenamefont {Sardina}\ \emph
  {et~al.}(2012{\natexlab{a}})\citenamefont {Sardina}, \citenamefont
  {Schlatter}, \citenamefont {Picano}, \citenamefont {Casciola}, \citenamefont
  {Brandt},\ and\ \citenamefont {Henningson}}]{sardina2012self}%
  \BibitemOpen
  \bibfield  {author} {\bibinfo {author} {\bibfnamefont {G.}~\bibnamefont
  {Sardina}}, \bibinfo {author} {\bibfnamefont {P.}~\bibnamefont {Schlatter}},
  \bibinfo {author} {\bibfnamefont {F.}~\bibnamefont {Picano}}, \bibinfo
  {author} {\bibfnamefont {C.~M.}\ \bibnamefont {Casciola}}, \bibinfo {author}
  {\bibfnamefont {L.}~\bibnamefont {Brandt}}, \ and\ \bibinfo {author}
  {\bibfnamefont {D.~S.}\ \bibnamefont {Henningson}},\ }\bibfield  {title}
  {\enquote {\bibinfo {title} {Self-similar transport of inertial particles in
  a turbulent boundary layer},}\ }\href@noop {} {\bibfield  {journal} {\bibinfo
   {journal} {Journal of Fluid Mechanics}\ }\textbf {\bibinfo {volume} {706}},\
  \bibinfo {pages} {584--596} (\bibinfo {year}
  {2012}{\natexlab{a}})}\BibitemShut {NoStop}%
\bibitem [{\citenamefont {Li}\ \emph {et~al.}(2016)\citenamefont {Li},
  \citenamefont {Wei}, \citenamefont {Luo},\ and\ \citenamefont
  {Fan}}]{li2016direct}%
  \BibitemOpen
  \bibfield  {author} {\bibinfo {author} {\bibfnamefont {D.}~\bibnamefont
  {Li}}, \bibinfo {author} {\bibfnamefont {A.}~\bibnamefont {Wei}}, \bibinfo
  {author} {\bibfnamefont {K.}~\bibnamefont {Luo}}, \ and\ \bibinfo {author}
  {\bibfnamefont {J.}~\bibnamefont {Fan}},\ }\bibfield  {title} {\enquote
  {\bibinfo {title} {Direct numerical simulation of a particle-laden flow in a
  flat plate boundary layer},}\ }\href@noop {} {\bibfield  {journal} {\bibinfo
  {journal} {International Journal of Multiphase Flow}\ }\textbf {\bibinfo
  {volume} {79}},\ \bibinfo {pages} {124--143} (\bibinfo {year}
  {2016})}\BibitemShut {NoStop}%
\bibitem [{\citenamefont {Bernardini}, \citenamefont {Pirozzoli},\ and\
  \citenamefont {Orlandi}(2013)}]{bernardini2013effect}%
  \BibitemOpen
  \bibfield  {author} {\bibinfo {author} {\bibfnamefont {M.}~\bibnamefont
  {Bernardini}}, \bibinfo {author} {\bibfnamefont {S.}~\bibnamefont
  {Pirozzoli}}, \ and\ \bibinfo {author} {\bibfnamefont {P.}~\bibnamefont
  {Orlandi}},\ }\bibfield  {title} {\enquote {\bibinfo {title} {The effect of
  large-scale turbulent structures on particle dispersion in wall-bounded
  flows},}\ }\href@noop {} {\bibfield  {journal} {\bibinfo  {journal}
  {International Journal of Multiphase Flow}\ }\textbf {\bibinfo {volume}
  {51}},\ \bibinfo {pages} {55--64} (\bibinfo {year} {2013})}\BibitemShut
  {NoStop}%
\bibitem [{\citenamefont {Marchioli}\ \emph {et~al.}(2008)\citenamefont
  {Marchioli}, \citenamefont {Soldati}, \citenamefont {Kuerten}, \citenamefont
  {Arcen}, \citenamefont {Taniere}, \citenamefont {Goldensoph}, \citenamefont
  {Squires}, \citenamefont {Cargnelutti},\ and\ \citenamefont
  {Portela}}]{marchioli2008statistics}%
  \BibitemOpen
  \bibfield  {author} {\bibinfo {author} {\bibfnamefont {C.}~\bibnamefont
  {Marchioli}}, \bibinfo {author} {\bibfnamefont {A.}~\bibnamefont {Soldati}},
  \bibinfo {author} {\bibfnamefont {J.~G.~M.}\ \bibnamefont {Kuerten}},
  \bibinfo {author} {\bibfnamefont {B.}~\bibnamefont {Arcen}}, \bibinfo
  {author} {\bibfnamefont {A.}~\bibnamefont {Taniere}}, \bibinfo {author}
  {\bibfnamefont {G.}~\bibnamefont {Goldensoph}}, \bibinfo {author}
  {\bibfnamefont {K.~D.}\ \bibnamefont {Squires}}, \bibinfo {author}
  {\bibfnamefont {M.~F.}\ \bibnamefont {Cargnelutti}}, \ and\ \bibinfo {author}
  {\bibfnamefont {L.~M.}\ \bibnamefont {Portela}},\ }\bibfield  {title}
  {\enquote {\bibinfo {title} {Statistics of particle dispersion in direct
  numerical simulations of wall-bounded turbulence: results of an international
  collaborative benchmark test},}\ }\href@noop {} {\bibfield  {journal}
  {\bibinfo  {journal} {International Journal of Multiphase Flow}\ }\textbf
  {\bibinfo {volume} {34}},\ \bibinfo {pages} {879--893} (\bibinfo {year}
  {2008})}\BibitemShut {NoStop}%
\bibitem [{\citenamefont {Marchioli}\ \emph {et~al.}(2003)\citenamefont
  {Marchioli}, \citenamefont {Giusti}, \citenamefont {Salvetti},\ and\
  \citenamefont {Soldati}}]{marchioli2003direct}%
  \BibitemOpen
  \bibfield  {author} {\bibinfo {author} {\bibfnamefont {C.}~\bibnamefont
  {Marchioli}}, \bibinfo {author} {\bibfnamefont {A.}~\bibnamefont {Giusti}},
  \bibinfo {author} {\bibfnamefont {M.~V.}\ \bibnamefont {Salvetti}}, \ and\
  \bibinfo {author} {\bibfnamefont {A.}~\bibnamefont {Soldati}},\ }\bibfield
  {title} {\enquote {\bibinfo {title} {Direct numerical simulation of particle
  wall transfer and deposition in upward turbulent pipe flow},}\ }\href@noop {}
  {\bibfield  {journal} {\bibinfo  {journal} {International journal of
  Multiphase flow}\ }\textbf {\bibinfo {volume} {29}},\ \bibinfo {pages}
  {1017--1038} (\bibinfo {year} {2003})}\BibitemShut {NoStop}%
\bibitem [{\citenamefont {Picano}, \citenamefont {Sardina},\ and\ \citenamefont
  {Casciola}(2009)}]{picano2009spatial}%
  \BibitemOpen
  \bibfield  {author} {\bibinfo {author} {\bibfnamefont {F.}~\bibnamefont
  {Picano}}, \bibinfo {author} {\bibfnamefont {G.}~\bibnamefont {Sardina}}, \
  and\ \bibinfo {author} {\bibfnamefont {C.~M.}\ \bibnamefont {Casciola}},\
  }\bibfield  {title} {\enquote {\bibinfo {title} {Spatial development of
  particle-laden turbulent pipe flow},}\ }\href@noop {} {\bibfield  {journal}
  {\bibinfo  {journal} {Physics of Fluids}\ }\textbf {\bibinfo {volume} {21}},\
  \bibinfo {pages} {093305} (\bibinfo {year} {2009})}\BibitemShut {NoStop}%
\bibitem [{\citenamefont {Sardina}\ \emph
  {et~al.}(2012{\natexlab{b}})\citenamefont {Sardina}, \citenamefont
  {Schlatter}, \citenamefont {Brandt}, \citenamefont {Picano},\ and\
  \citenamefont {Casciola}}]{sardina2012wall}%
  \BibitemOpen
  \bibfield  {author} {\bibinfo {author} {\bibfnamefont {G.}~\bibnamefont
  {Sardina}}, \bibinfo {author} {\bibfnamefont {P.}~\bibnamefont {Schlatter}},
  \bibinfo {author} {\bibfnamefont {L.}~\bibnamefont {Brandt}}, \bibinfo
  {author} {\bibfnamefont {F.}~\bibnamefont {Picano}}, \ and\ \bibinfo {author}
  {\bibfnamefont {C.~M.}\ \bibnamefont {Casciola}},\ }\bibfield  {title}
  {\enquote {\bibinfo {title} {Wall accumulation and spatial localization in
  particle-laden wall flows},}\ }\href@noop {} {\bibfield  {journal} {\bibinfo
  {journal} {Journal of Fluid Mechanics}\ }\textbf {\bibinfo {volume} {699}},\
  \bibinfo {pages} {50--78} (\bibinfo {year} {2012}{\natexlab{b}})}\BibitemShut
  {NoStop}%
\bibitem [{\citenamefont {Kulick}, \citenamefont {Fessler},\ and\ \citenamefont
  {Eaton}(1994)}]{kulick1994particle}%
  \BibitemOpen
  \bibfield  {author} {\bibinfo {author} {\bibfnamefont {J.~D.}\ \bibnamefont
  {Kulick}}, \bibinfo {author} {\bibfnamefont {J.~R.}\ \bibnamefont {Fessler}},
  \ and\ \bibinfo {author} {\bibfnamefont {J.~K.}\ \bibnamefont {Eaton}},\
  }\bibfield  {title} {\enquote {\bibinfo {title} {Particle response and
  turbulence modification in fully developed channel flow},}\ }\href@noop {}
  {\bibfield  {journal} {\bibinfo  {journal} {Journal of Fluid Mechanics}\
  }\textbf {\bibinfo {volume} {277}},\ \bibinfo {pages} {109--134} (\bibinfo
  {year} {1994})}\BibitemShut {NoStop}%
\bibitem [{\citenamefont {Liu}, \citenamefont {Luo},\ and\ \citenamefont
  {Fan}(2016)}]{liu2016turbulence}%
  \BibitemOpen
  \bibfield  {author} {\bibinfo {author} {\bibfnamefont {X.}~\bibnamefont
  {Liu}}, \bibinfo {author} {\bibfnamefont {K.}~\bibnamefont {Luo}}, \ and\
  \bibinfo {author} {\bibfnamefont {J.}~\bibnamefont {Fan}},\ }\bibfield
  {title} {\enquote {\bibinfo {title} {Turbulence modulation in a
  particle-laden flow over a hemisphere-roughened wall},}\ }\href@noop {}
  {\bibfield  {journal} {\bibinfo  {journal} {International Journal of
  Multiphase Flow}\ }\textbf {\bibinfo {volume} {87}},\ \bibinfo {pages}
  {250--262} (\bibinfo {year} {2016})}\BibitemShut {NoStop}%
\bibitem [{\citenamefont {De~Marchis}\ \emph {et~al.}(2016)\citenamefont
  {De~Marchis}, \citenamefont {Milici}, \citenamefont {Sardina},\ and\
  \citenamefont {Napoli}}]{de2016interaction}%
  \BibitemOpen
  \bibfield  {author} {\bibinfo {author} {\bibfnamefont {M.}~\bibnamefont
  {De~Marchis}}, \bibinfo {author} {\bibfnamefont {B.}~\bibnamefont {Milici}},
  \bibinfo {author} {\bibfnamefont {G.}~\bibnamefont {Sardina}}, \ and\
  \bibinfo {author} {\bibfnamefont {E.}~\bibnamefont {Napoli}},\ }\bibfield
  {title} {\enquote {\bibinfo {title} {Interaction between turbulent structures
  and particles in roughened channel},}\ }\href@noop {} {\bibfield  {journal}
  {\bibinfo  {journal} {International Journal of Multiphase Flow}\ }\textbf
  {\bibinfo {volume} {78}},\ \bibinfo {pages} {117--131} (\bibinfo {year}
  {2016})}\BibitemShut {NoStop}%
\bibitem [{\citenamefont {Vreman}(2015)}]{vreman2015turbulence}%
  \BibitemOpen
  \bibfield  {author} {\bibinfo {author} {\bibfnamefont {A.~W.}\ \bibnamefont
  {Vreman}},\ }\bibfield  {title} {\enquote {\bibinfo {title} {Turbulence
  attenuation in particle-laden flow in smooth and rough channels},}\
  }\href@noop {} {\bibfield  {journal} {\bibinfo  {journal} {Journal of Fluid
  Mechanics}\ }\textbf {\bibinfo {volume} {773}},\ \bibinfo {pages} {103}
  (\bibinfo {year} {2015})}\BibitemShut {NoStop}%
\bibitem [{\citenamefont {Picano}\ \emph {et~al.}(2011)\citenamefont {Picano},
  \citenamefont {Battista}, \citenamefont {Troiani},\ and\ \citenamefont
  {Casciola}}]{picano2011dynamics}%
  \BibitemOpen
  \bibfield  {author} {\bibinfo {author} {\bibfnamefont {F.}~\bibnamefont
  {Picano}}, \bibinfo {author} {\bibfnamefont {F.}~\bibnamefont {Battista}},
  \bibinfo {author} {\bibfnamefont {G.}~\bibnamefont {Troiani}}, \ and\
  \bibinfo {author} {\bibfnamefont {C.~M.}\ \bibnamefont {Casciola}},\
  }\bibfield  {title} {\enquote {\bibinfo {title} {Dynamics of piv seeding
  particles in turbulent premixed flames},}\ }\href@noop {} {\bibfield
  {journal} {\bibinfo  {journal} {Experiments in Fluids}\ }\textbf {\bibinfo
  {volume} {50}},\ \bibinfo {pages} {75--88} (\bibinfo {year}
  {2011})}\BibitemShut {NoStop}%
\bibitem [{\citenamefont {Gualtieri}, \citenamefont {Battista},\ and\
  \citenamefont {Casciola}(2017)}]{gualtieri2017turbulence}%
  \BibitemOpen
  \bibfield  {author} {\bibinfo {author} {\bibfnamefont {P.}~\bibnamefont
  {Gualtieri}}, \bibinfo {author} {\bibfnamefont {F.}~\bibnamefont {Battista}},
  \ and\ \bibinfo {author} {\bibfnamefont {C.}~\bibnamefont {Casciola}},\
  }\bibfield  {title} {\enquote {\bibinfo {title} {Turbulence modulation in
  heavy-loaded suspensions of tiny particles},}\ }\href@noop {} {\bibfield
  {journal} {\bibinfo  {journal} {Physical Review Fluids}\ }\textbf {\bibinfo
  {volume} {2}},\ \bibinfo {pages} {034304} (\bibinfo {year}
  {2017})}\BibitemShut {NoStop}%
\bibitem [{\citenamefont {Battista}\ \emph {et~al.}(2011)\citenamefont
  {Battista}, \citenamefont {Picano}, \citenamefont {Troiani},\ and\
  \citenamefont {Casciola}}]{battista2011intermittent}%
  \BibitemOpen
  \bibfield  {author} {\bibinfo {author} {\bibfnamefont {F.}~\bibnamefont
  {Battista}}, \bibinfo {author} {\bibfnamefont {F.}~\bibnamefont {Picano}},
  \bibinfo {author} {\bibfnamefont {G.}~\bibnamefont {Troiani}}, \ and\
  \bibinfo {author} {\bibfnamefont {C.~M.}\ \bibnamefont {Casciola}},\
  }\bibfield  {title} {\enquote {\bibinfo {title} {Intermittent features of
  inertial particle distributions in turbulent premixed flames},}\ }\href@noop
  {} {\bibfield  {journal} {\bibinfo  {journal} {Physics of Fluids}\ }\textbf
  {\bibinfo {volume} {23}},\ \bibinfo {pages} {123304} (\bibinfo {year}
  {2011})}\BibitemShut {NoStop}%
\bibitem [{\citenamefont {Wu}\ \emph {et~al.}(2017)\citenamefont {Wu},
  \citenamefont {Soligo}, \citenamefont {Marchioli}, \citenamefont {Soldati},\
  and\ \citenamefont {Piomelli}}]{wu2017particle}%
  \BibitemOpen
  \bibfield  {author} {\bibinfo {author} {\bibfnamefont {W.}~\bibnamefont
  {Wu}}, \bibinfo {author} {\bibfnamefont {G.~G.}\ \bibnamefont {Soligo}},
  \bibinfo {author} {\bibfnamefont {C.}~\bibnamefont {Marchioli}}, \bibinfo
  {author} {\bibfnamefont {A.}~\bibnamefont {Soldati}}, \ and\ \bibinfo
  {author} {\bibfnamefont {U.}~\bibnamefont {Piomelli}},\ }\bibfield  {title}
  {\enquote {\bibinfo {title} {Particle resuspension by a periodically forced
  impinging jet},}\ }\href@noop {} {\bibfield  {journal} {\bibinfo  {journal}
  {Journal of Fluid Mechanics}\ }\textbf {\bibinfo {volume} {820}},\ \bibinfo
  {pages} {284--311} (\bibinfo {year} {2017})}\BibitemShut {NoStop}%
\bibitem [{\citenamefont {Lau}\ and\ \citenamefont
  {Nathan}(2014)}]{lau2014influence}%
  \BibitemOpen
  \bibfield  {author} {\bibinfo {author} {\bibfnamefont {T.~C.~W.}\
  \bibnamefont {Lau}}\ and\ \bibinfo {author} {\bibfnamefont {G.~J.}\
  \bibnamefont {Nathan}},\ }\bibfield  {title} {\enquote {\bibinfo {title}
  {Influence of stokes number on the velocity and concentration distributions
  in particle-laden jets},}\ }\href@noop {} {\bibfield  {journal} {\bibinfo
  {journal} {Journal of Fluid Mechanics}\ }\textbf {\bibinfo {volume} {757}},\
  \bibinfo {pages} {432--457} (\bibinfo {year} {2014})}\BibitemShut {NoStop}%
\bibitem [{\citenamefont {Lau}\ and\ \citenamefont
  {Nathan}(2016)}]{lau2016effect}%
  \BibitemOpen
  \bibfield  {author} {\bibinfo {author} {\bibfnamefont {T.~C.~W.}\
  \bibnamefont {Lau}}\ and\ \bibinfo {author} {\bibfnamefont {G.~J.}\
  \bibnamefont {Nathan}},\ }\bibfield  {title} {\enquote {\bibinfo {title} {The
  effect of stokes number on particle velocity and concentration distributions
  in a well-characterised, turbulent, co-flowing two-phase jet},}\ }\href@noop
  {} {\bibfield  {journal} {\bibinfo  {journal} {Journal of Fluid Mechanics}\
  }\textbf {\bibinfo {volume} {809}},\ \bibinfo {pages} {72--110} (\bibinfo
  {year} {2016})}\BibitemShut {NoStop}%
\bibitem [{\citenamefont {Wang}, \citenamefont {Zheng},\ and\ \citenamefont
  {Wang}(2017)}]{wang2017direct}%
  \BibitemOpen
  \bibfield  {author} {\bibinfo {author} {\bibfnamefont {X.}~\bibnamefont
  {Wang}}, \bibinfo {author} {\bibfnamefont {X.}~\bibnamefont {Zheng}}, \ and\
  \bibinfo {author} {\bibfnamefont {P.}~\bibnamefont {Wang}},\ }\bibfield
  {title} {\enquote {\bibinfo {title} {Direct numerical simulation of
  particle-laden plane turbulent wall jet and the influence of stokes
  number},}\ }\href@noop {} {\bibfield  {journal} {\bibinfo  {journal}
  {International Journal of Multiphase Flow}\ }\textbf {\bibinfo {volume}
  {92}},\ \bibinfo {pages} {82--92} (\bibinfo {year} {2017})}\BibitemShut
  {NoStop}%
\bibitem [{\citenamefont {Abdelaziz}\ \emph {et~al.}(2017)\citenamefont
  {Abdelaziz}, \citenamefont {Gaber}, \citenamefont {Abd-Elwakil},
  \citenamefont {Mabrouk}, \citenamefont {Elgohary}, \citenamefont {Kamel},
  \citenamefont {Kabary}, \citenamefont {Freag}, \citenamefont {Samaha},
  \citenamefont {Mortada} \emph {et~al.}}]{abdelaziz2017inhalable}%
  \BibitemOpen
  \bibfield  {author} {\bibinfo {author} {\bibfnamefont {H.~M.}\ \bibnamefont
  {Abdelaziz}}, \bibinfo {author} {\bibfnamefont {M.}~\bibnamefont {Gaber}},
  \bibinfo {author} {\bibfnamefont {M.~M.}\ \bibnamefont {Abd-Elwakil}},
  \bibinfo {author} {\bibfnamefont {M.~T.}\ \bibnamefont {Mabrouk}}, \bibinfo
  {author} {\bibfnamefont {M.~M.}\ \bibnamefont {Elgohary}}, \bibinfo {author}
  {\bibfnamefont {N.~M.}\ \bibnamefont {Kamel}}, \bibinfo {author}
  {\bibfnamefont {D.~M.}\ \bibnamefont {Kabary}}, \bibinfo {author}
  {\bibfnamefont {M.~S.}\ \bibnamefont {Freag}}, \bibinfo {author}
  {\bibfnamefont {M.~W.}\ \bibnamefont {Samaha}}, \bibinfo {author}
  {\bibfnamefont {S.~M.}\ \bibnamefont {Mortada}},  \emph {et~al.},\ }\bibfield
   {title} {\enquote {\bibinfo {title} {Inhalable particulate drug delivery
  systems for lung cancer therapy: Nanoparticles, microparticles,
  nanocomposites and nanoaggregates},}\ }\href@noop {} {\bibfield  {journal}
  {\bibinfo  {journal} {Journal of Controlled Release}\ } (\bibinfo {year}
  {2017})}\BibitemShut {NoStop}%
\bibitem [{\citenamefont {Ni}\ \emph {et~al.}(2017)\citenamefont {Ni},
  \citenamefont {Zhao}, \citenamefont {Liu}, \citenamefont {Liang},
  \citenamefont {Muenster},\ and\ \citenamefont {Mao}}]{ni2017nanocrystals}%
  \BibitemOpen
  \bibfield  {author} {\bibinfo {author} {\bibfnamefont {R.}~\bibnamefont
  {Ni}}, \bibinfo {author} {\bibfnamefont {J.}~\bibnamefont {Zhao}}, \bibinfo
  {author} {\bibfnamefont {Q.}~\bibnamefont {Liu}}, \bibinfo {author}
  {\bibfnamefont {Z.}~\bibnamefont {Liang}}, \bibinfo {author} {\bibfnamefont
  {U.}~\bibnamefont {Muenster}}, \ and\ \bibinfo {author} {\bibfnamefont
  {S.}~\bibnamefont {Mao}},\ }\bibfield  {title} {\enquote {\bibinfo {title}
  {Nanocrystals embedded in chitosan-based respirable swellable microparticles
  as dry powder for sustained pulmonary drug delivery},}\ }\href@noop {}
  {\bibfield  {journal} {\bibinfo  {journal} {European Journal of
  Pharmaceutical Sciences}\ }\textbf {\bibinfo {volume} {99}},\ \bibinfo
  {pages} {137--146} (\bibinfo {year} {2017})}\BibitemShut {NoStop}%
\bibitem [{\citenamefont {Ghahramani}\ \emph {et~al.}(2017)\citenamefont
  {Ghahramani}, \citenamefont {Abouali}, \citenamefont {Emdad},\ and\
  \citenamefont {Ahmadi}}]{ghahramani2017numerical}%
  \BibitemOpen
  \bibfield  {author} {\bibinfo {author} {\bibfnamefont {E.}~\bibnamefont
  {Ghahramani}}, \bibinfo {author} {\bibfnamefont {O.}~\bibnamefont {Abouali}},
  \bibinfo {author} {\bibfnamefont {H.}~\bibnamefont {Emdad}}, \ and\ \bibinfo
  {author} {\bibfnamefont {G.}~\bibnamefont {Ahmadi}},\ }\bibfield  {title}
  {\enquote {\bibinfo {title} {Numerical investigation of turbulent airflow and
  microparticle deposition in a realistic model of human upper airway using
  les},}\ }\href@noop {} {\bibfield  {journal} {\bibinfo  {journal} {Computers
  \& Fluids}\ }\textbf {\bibinfo {volume} {157}},\ \bibinfo {pages} {43--54}
  (\bibinfo {year} {2017})}\BibitemShut {NoStop}%
\bibitem [{\citenamefont {Stylianou}, \citenamefont {Sznitman},\ and\
  \citenamefont {Kassinos}(2016)}]{stylianou2016direct}%
  \BibitemOpen
  \bibfield  {author} {\bibinfo {author} {\bibfnamefont {F.~S.}\ \bibnamefont
  {Stylianou}}, \bibinfo {author} {\bibfnamefont {J.}~\bibnamefont {Sznitman}},
  \ and\ \bibinfo {author} {\bibfnamefont {S.~C.}\ \bibnamefont {Kassinos}},\
  }\bibfield  {title} {\enquote {\bibinfo {title} {Direct numerical simulation
  of particle laden flow in a human airway bifurcation model},}\ }\href@noop {}
  {\bibfield  {journal} {\bibinfo  {journal} {International Journal of Heat and
  Fluid Flow}\ }\textbf {\bibinfo {volume} {61}},\ \bibinfo {pages} {677--710}
  (\bibinfo {year} {2016})}\BibitemShut {NoStop}%
\bibitem [{\citenamefont {Rahimi-Gorji}\ \emph {et~al.}(2015)\citenamefont
  {Rahimi-Gorji}, \citenamefont {Pourmehran}, \citenamefont {Gorji-Bandpy},\
  and\ \citenamefont {Gorji}}]{rahimi2015cfd}%
  \BibitemOpen
  \bibfield  {author} {\bibinfo {author} {\bibfnamefont {M.}~\bibnamefont
  {Rahimi-Gorji}}, \bibinfo {author} {\bibfnamefont {O.}~\bibnamefont
  {Pourmehran}}, \bibinfo {author} {\bibfnamefont {M.}~\bibnamefont
  {Gorji-Bandpy}}, \ and\ \bibinfo {author} {\bibfnamefont {T.~B.}\
  \bibnamefont {Gorji}},\ }\bibfield  {title} {\enquote {\bibinfo {title} {Cfd
  simulation of airflow behavior and particle transport and deposition in
  different breathing conditions through the realistic model of human
  airways},}\ }\href@noop {} {\bibfield  {journal} {\bibinfo  {journal}
  {Journal of Molecular Liquids}\ }\textbf {\bibinfo {volume} {209}},\ \bibinfo
  {pages} {121--133} (\bibinfo {year} {2015})}\BibitemShut {NoStop}%
\bibitem [{\citenamefont {Thondapu}\ \emph {et~al.}(2016)\citenamefont
  {Thondapu}, \citenamefont {Bourantas}, \citenamefont {Foin}, \citenamefont
  {Jang}, \citenamefont {Serruys},\ and\ \citenamefont
  {Barlis}}]{thondapu2016biomechanical}%
  \BibitemOpen
  \bibfield  {author} {\bibinfo {author} {\bibfnamefont {V.}~\bibnamefont
  {Thondapu}}, \bibinfo {author} {\bibfnamefont {C.~V.}\ \bibnamefont
  {Bourantas}}, \bibinfo {author} {\bibfnamefont {N.}~\bibnamefont {Foin}},
  \bibinfo {author} {\bibfnamefont {I.-K.}\ \bibnamefont {Jang}}, \bibinfo
  {author} {\bibfnamefont {P.~W.}\ \bibnamefont {Serruys}}, \ and\ \bibinfo
  {author} {\bibfnamefont {P.}~\bibnamefont {Barlis}},\ }\bibfield  {title}
  {\enquote {\bibinfo {title} {Biomechanical stress in coronary
  atherosclerosis: emerging insights from computational modelling},}\
  }\href@noop {} {\bibfield  {journal} {\bibinfo  {journal} {European heart
  journal}\ }\textbf {\bibinfo {volume} {38}},\ \bibinfo {pages} {81--92}
  (\bibinfo {year} {2016})}\BibitemShut {NoStop}%
\bibitem [{\citenamefont {Choi}\ \emph {et~al.}(2018)\citenamefont {Choi},
  \citenamefont {Park}, \citenamefont {Byeon},\ and\ \citenamefont
  {Lee}}]{choi2018flow}%
  \BibitemOpen
  \bibfield  {author} {\bibinfo {author} {\bibfnamefont {W.}~\bibnamefont
  {Choi}}, \bibinfo {author} {\bibfnamefont {J.~H.}\ \bibnamefont {Park}},
  \bibinfo {author} {\bibfnamefont {H.}~\bibnamefont {Byeon}}, \ and\ \bibinfo
  {author} {\bibfnamefont {S.~J.}\ \bibnamefont {Lee}},\ }\bibfield  {title}
  {\enquote {\bibinfo {title} {Flow characteristics around a deformable
  stenosis under pulsatile flow condition},}\ }\href@noop {} {\bibfield
  {journal} {\bibinfo  {journal} {Physics of Fluids}\ }\textbf {\bibinfo
  {volume} {30}},\ \bibinfo {pages} {011902} (\bibinfo {year}
  {2018})}\BibitemShut {NoStop}%
\bibitem [{\citenamefont {Huang}\ and\ \citenamefont
  {Durbin}(2010)}]{huang_2010}%
  \BibitemOpen
  \bibfield  {author} {\bibinfo {author} {\bibfnamefont {X.}~\bibnamefont
  {Huang}}\ and\ \bibinfo {author} {\bibfnamefont {P.}~\bibnamefont {Durbin}},\
  }\bibfield  {title} {\enquote {\bibinfo {title} {Particulate dispersion in a
  turbulent serpentine channel},}\ }\href@noop {} {\bibfield  {journal}
  {\bibinfo  {journal} {Flow, turbulence and combustion}\ }\textbf {\bibinfo
  {volume} {85}},\ \bibinfo {pages} {333--344} (\bibinfo {year}
  {2010})}\BibitemShut {NoStop}%
\bibitem [{\citenamefont {Huang}\ and\ \citenamefont
  {Durbin}(2012)}]{huang_2012}%
  \BibitemOpen
  \bibfield  {author} {\bibinfo {author} {\bibfnamefont {X.}~\bibnamefont
  {Huang}}\ and\ \bibinfo {author} {\bibfnamefont {P.~A.}\ \bibnamefont
  {Durbin}},\ }\bibfield  {title} {\enquote {\bibinfo {title} {Particulate
  mixing in a turbulent serpentine duct},}\ }\href@noop {} {\bibfield
  {journal} {\bibinfo  {journal} {Physics of Fluids}\ }\textbf {\bibinfo
  {volume} {24}},\ \bibinfo {pages} {013301} (\bibinfo {year}
  {2012})}\BibitemShut {NoStop}%
\bibitem [{\citenamefont {Noorani}\ \emph {et~al.}(2015)\citenamefont
  {Noorani}, \citenamefont {Sardina}, \citenamefont {Brandt},\ and\
  \citenamefont {Schlatter}}]{noorani_2015}%
  \BibitemOpen
  \bibfield  {author} {\bibinfo {author} {\bibfnamefont {A.}~\bibnamefont
  {Noorani}}, \bibinfo {author} {\bibfnamefont {G.}~\bibnamefont {Sardina}},
  \bibinfo {author} {\bibfnamefont {L.}~\bibnamefont {Brandt}}, \ and\ \bibinfo
  {author} {\bibfnamefont {P.}~\bibnamefont {Schlatter}},\ }\bibfield  {title}
  {\enquote {\bibinfo {title} {Particle velocity and acceleration in turbulent
  bent pipe flows},}\ }\href@noop {} {\bibfield  {journal} {\bibinfo  {journal}
  {Flow, Turbulence and Combustion}\ }\textbf {\bibinfo {volume} {95}},\
  \bibinfo {pages} {539--559} (\bibinfo {year} {2015})}\BibitemShut {NoStop}%
\bibitem [{\citenamefont {Ault}\ \emph {et~al.}(2016)\citenamefont {Ault},
  \citenamefont {Fani}, \citenamefont {Chen}, \citenamefont {Shin},
  \citenamefont {Gallaire},\ and\ \citenamefont {Stone}}]{ault2016vortex}%
  \BibitemOpen
  \bibfield  {author} {\bibinfo {author} {\bibfnamefont {J.~T.}\ \bibnamefont
  {Ault}}, \bibinfo {author} {\bibfnamefont {A.}~\bibnamefont {Fani}}, \bibinfo
  {author} {\bibfnamefont {K.~K.}\ \bibnamefont {Chen}}, \bibinfo {author}
  {\bibfnamefont {S.}~\bibnamefont {Shin}}, \bibinfo {author} {\bibfnamefont
  {F.}~\bibnamefont {Gallaire}}, \ and\ \bibinfo {author} {\bibfnamefont
  {H.~A.}\ \bibnamefont {Stone}},\ }\bibfield  {title} {\enquote {\bibinfo
  {title} {Vortex-breakdown-induced particle capture in branching junctions},}\
  }\href@noop {} {\bibfield  {journal} {\bibinfo  {journal} {Physical review
  letters}\ }\textbf {\bibinfo {volume} {117}},\ \bibinfo {pages} {084501}
  (\bibinfo {year} {2016})}\BibitemShut {NoStop}%
\bibitem [{\citenamefont {Stella}, \citenamefont {Mazellier},\ and\
  \citenamefont {Kourta}(2017)}]{stella2017scaling}%
  \BibitemOpen
  \bibfield  {author} {\bibinfo {author} {\bibfnamefont {F.}~\bibnamefont
  {Stella}}, \bibinfo {author} {\bibfnamefont {N.}~\bibnamefont {Mazellier}}, \
  and\ \bibinfo {author} {\bibfnamefont {A.}~\bibnamefont {Kourta}},\
  }\bibfield  {title} {\enquote {\bibinfo {title} {Scaling of separated shear
  layers: an investigation of mass entrainment},}\ }\href@noop {} {\bibfield
  {journal} {\bibinfo  {journal} {Journal of Fluid Mechanics}\ }\textbf
  {\bibinfo {volume} {826}},\ \bibinfo {pages} {851--887} (\bibinfo {year}
  {2017})}\BibitemShut {NoStop}%
\bibitem [{\citenamefont {Krank}, \citenamefont {Kronbichler},\ and\
  \citenamefont {Wall}(2017)}]{krankdirect}%
  \BibitemOpen
  \bibfield  {author} {\bibinfo {author} {\bibfnamefont {B.}~\bibnamefont
  {Krank}}, \bibinfo {author} {\bibfnamefont {M.}~\bibnamefont {Kronbichler}},
  \ and\ \bibinfo {author} {\bibfnamefont {W.~A.}\ \bibnamefont {Wall}},\
  }\bibfield  {title} {\enquote {\bibinfo {title} {Direct numerical simulation
  of flow over periodic hills up to reh=10595},}\ }\href@noop {} {\bibfield
  {journal} {\bibinfo  {journal} {Flow, Turbulence and Combustion}\ ,\ \bibinfo
  {pages} {1--31}} (\bibinfo {year} {2017})}\BibitemShut {NoStop}%
\bibitem [{\citenamefont {Schiavo}, \citenamefont {Wolf},\ and\ \citenamefont
  {Azevedo}(2017)}]{schiavo2017turbulent}%
  \BibitemOpen
  \bibfield  {author} {\bibinfo {author} {\bibfnamefont {L.~A. C.~A.}\
  \bibnamefont {Schiavo}}, \bibinfo {author} {\bibfnamefont {W.~R.}\
  \bibnamefont {Wolf}}, \ and\ \bibinfo {author} {\bibfnamefont {J.~L.~F.}\
  \bibnamefont {Azevedo}},\ }\bibfield  {title} {\enquote {\bibinfo {title}
  {Turbulent kinetic energy budgets in wall bounded flows with pressure
  gradients and separation},}\ }\href@noop {} {\bibfield  {journal} {\bibinfo
  {journal} {Physics of Fluids}\ }\textbf {\bibinfo {volume} {29}},\ \bibinfo
  {pages} {115108} (\bibinfo {year} {2017})}\BibitemShut {NoStop}%
\bibitem [{\citenamefont {Mollicone}\ \emph {et~al.}(2018)\citenamefont
  {Mollicone}, \citenamefont {Battista}, \citenamefont {Gualtieri},\ and\
  \citenamefont {Casciola}}]{mollicone2018turbulence}%
  \BibitemOpen
  \bibfield  {author} {\bibinfo {author} {\bibfnamefont {J.-P.}\ \bibnamefont
  {Mollicone}}, \bibinfo {author} {\bibfnamefont {F.}~\bibnamefont {Battista}},
  \bibinfo {author} {\bibfnamefont {P.}~\bibnamefont {Gualtieri}}, \ and\
  \bibinfo {author} {\bibfnamefont {C.~M.}\ \bibnamefont {Casciola}},\
  }\bibfield  {title} {\enquote {\bibinfo {title} {Turbulence dynamics in
  separated flows: the generalised kolmogorov equation for inhomogeneous
  anisotropic conditions},}\ }\href@noop {} {\bibfield  {journal} {\bibinfo
  {journal} {Journal of Fluid Mechanics}\ }\textbf {\bibinfo {volume} {841}},\
  \bibinfo {pages} {1012--1039} (\bibinfo {year} {2018})}\BibitemShut {NoStop}%
\bibitem [{\citenamefont {Mollicone}\ \emph {et~al.}(2017)\citenamefont
  {Mollicone}, \citenamefont {Battista}, \citenamefont {Gualtieri},\ and\
  \citenamefont {Casciola}}]{Mollicone_2017}%
  \BibitemOpen
  \bibfield  {author} {\bibinfo {author} {\bibfnamefont {J.-P.}\ \bibnamefont
  {Mollicone}}, \bibinfo {author} {\bibfnamefont {F.}~\bibnamefont {Battista}},
  \bibinfo {author} {\bibfnamefont {P.}~\bibnamefont {Gualtieri}}, \ and\
  \bibinfo {author} {\bibfnamefont {C.~M.}\ \bibnamefont {Casciola}},\
  }\bibfield  {title} {\enquote {\bibinfo {title} {Effect of geometry and
  reynolds number on the turbulent separated flow behind a bulge in a
  channel},}\ }\href@noop {} {\bibfield  {journal} {\bibinfo  {journal}
  {Journal of Fluid Mechanics}\ }\textbf {\bibinfo {volume} {823}},\ \bibinfo
  {pages} {100--133} (\bibinfo {year} {2017})}\BibitemShut {NoStop}%
\bibitem [{\citenamefont {Passaggia}\ and\ \citenamefont
  {Ehrenstein}(2018)}]{passaggia2018optimal}%
  \BibitemOpen
  \bibfield  {author} {\bibinfo {author} {\bibfnamefont {P.-Y.}\ \bibnamefont
  {Passaggia}}\ and\ \bibinfo {author} {\bibfnamefont {U.}~\bibnamefont
  {Ehrenstein}},\ }\bibfield  {title} {\enquote {\bibinfo {title} {Optimal
  control of a separated boundary-layer flow over a bump},}\ }\href@noop {}
  {\bibfield  {journal} {\bibinfo  {journal} {Journal of Fluid Mechanics}\
  }\textbf {\bibinfo {volume} {840}},\ \bibinfo {pages} {238--265} (\bibinfo
  {year} {2018})}\BibitemShut {NoStop}%
\bibitem [{\citenamefont {K{\"{a}}hler}, \citenamefont {Scharnowski},\ and\
  \citenamefont {Cierpka}(2016)}]{Kahler_2016}%
  \BibitemOpen
  \bibfield  {author} {\bibinfo {author} {\bibfnamefont {C.}~\bibnamefont
  {K{\"{a}}hler}}, \bibinfo {author} {\bibfnamefont {S.}~\bibnamefont
  {Scharnowski}}, \ and\ \bibinfo {author} {\bibfnamefont {C.}~\bibnamefont
  {Cierpka}},\ }\bibfield  {title} {\enquote {\bibinfo {title} {{Highly
  resolved experimental results of the separated flow in a channel with
  streamwise periodic constrictions}},}\ }\href@noop {} {\bibfield  {journal}
  {\bibinfo  {journal} {Journal of Fluid Mechanics}\ }\textbf {\bibinfo
  {volume} {796}},\ \bibinfo {pages} {257--284} (\bibinfo {year}
  {2016})}\BibitemShut {NoStop}%
\bibitem [{\citenamefont {Fischer}, \citenamefont {Lottes},\ and\ \citenamefont
  {Kerkemeier}(2008)}]{nek5000}%
  \BibitemOpen
  \bibfield  {author} {\bibinfo {author} {\bibfnamefont {P.}~\bibnamefont
  {Fischer}}, \bibinfo {author} {\bibfnamefont {J.~W.}\ \bibnamefont {Lottes}},
  \ and\ \bibinfo {author} {\bibfnamefont {S.~G.}\ \bibnamefont {Kerkemeier}},\
  }\href@noop {} {\enquote {\bibinfo {title} {Nek5000 - {O}pen source spectral
  element {CFD} solver. {A}rgonne {N}ational {L}aboratory, {M}athematics and
  {C}omputer {S}cience {D}ivision, {A}rgonne, {IL}, see
  http://nek5000.mcs.anl.gov},}\ } (\bibinfo {year} {2008})\BibitemShut
  {NoStop}%
\bibitem [{\citenamefont {Patera}(1984)}]{Patera_1984}%
  \BibitemOpen
  \bibfield  {author} {\bibinfo {author} {\bibfnamefont {A.~T.}\ \bibnamefont
  {Patera}},\ }\bibfield  {title} {\enquote {\bibinfo {title} {A spectral
  element method for fluid dynamics},}\ }\href@noop {} {\bibfield  {journal}
  {\bibinfo  {journal} {Journal of Computational Physics}\ }\textbf {\bibinfo
  {volume} {54}},\ \bibinfo {pages} {468--488} (\bibinfo {year}
  {1984})}\BibitemShut {NoStop}%
\end{thebibliography}%

\end{document}